\documentclass[11pt]{article}
\usepackage{amsmath,amsfonts, epsf,epsfig,color,latexsym,mathrsfs}
\usepackage{amssymb,graphicx}
\numberwithin{equation}{section}
\usepackage{bbm}

\usepackage[utf8]{inputenc}
\usepackage[T1]{fontenc}

\usepackage{lmodern}

\def\Z {\mathbb{Z}}

\def\C {\mathbb{C}}

\def\d{\text{d}}
\def\G{\Gamma}

\begin{document}

\thispagestyle{empty}
\setlength\textwidth{450pt}

\def\thefootnote{\fnsymbol{footnote}}\begin{flushright}
\end{flushright}\vskip 0.5cm\begin{center}
\Large{\bf  On Type IIA geometries dual to ${\cal N}=2$ SCFTs}
\end{center}\vskip 0.8cm
\begin{center}{\large R. A.  Reid-Edwards$^{1}$, B. Stefa\'nski, jr.$^{2}$}
\vskip 0.2cm{\it $^1$ Mathematical Institute, Oxford University \\ 24--29 St Giles' 
Oxford OX1 3LB, UK }
\vskip 0.2cm{\it $^2$ Centre for Mathematical Science, City University London,
\\ Northampton Square, London EC1V 0HB, UK}
\end{center}
\vskip 1.0cm
\begin{abstract}\noindent
We provide explicit solutions of Type IIA supergravity which are believed to be dual to ${\cal N}=2$ superconformal four dimensional gauge theories. These explicit solutions are based on the general ansatz for such a type of backgrounds introduced by Gaiotto and Maldacena.
\end{abstract}

\vfill

\setcounter{footnote}{0}
\def\thefootnote{\arabic{footnote}}
\newpage

\renewcommand{\theequation}{\thesection.\arabic{equation}}

\section{Introduction}
\label{sec0}

The gauge/string correspondence~\cite{Maldacena:1997re} has provided us with a fundamentally new way of investigating strongly coupled phenomena. In its simplest formulation, it asserts that strongly coupled systems, such as four dimensional gauge theories, have a dual description as a string- or M-theory on a curved spacetime. While it is not clear whether some form of the gauge/string correspondence will eventually apply to all strongly coupled (field-theory) systems, large classes of dual theories have been identified. The best-understood dual pairs involve $d$ dimensional super-conformal field theories (SCFTs), whose duals are described by string/M-theory on supersymmetric $AdS_{d+1}\times X$ spacetimes where $X$ is a $D_c-d-1$-dimensional 
manifold.~\footnote{$D_c$ is the critical dimension of string- or M-theory: $D_c=10$ or $11$.} 

For each $d$ there are maximally supersymmetric dual pairs such as the celebrated ${\cal N}=4$ super-Yang-Mills (SYM) $SU(N)$ gauge theory and the dual Type IIB string theory on $AdS_5\times S^5$ or the more recently found~\cite{Aharony:2008ug} super-Chern-Simons theory coupled to matter and its M-theory dual $AdS_4\times (S^7/\mathbbm{Z}_k)$. In these and related examples integrability techniques have provided a very detailed understanding of anomalous dimensions of operators, or the spectrum of the dual string theory, at all values (weak, strong and intermediate) of the gauge coupling.~\footnote{As suggested by 't Hooft~\cite{'tHooft:1973jz}, these results are limited to the large $N$, or planar,  limit.} These explicit expressions for generic gauge theory quantities which are not protected by non-renormalisation theorems should be taken as very strong evidence for the validity of the gauge/string duality for these dual pairs. More importantly though, it gives one confidence that the gauge/string approach captures some of the underlying fundamental features of wider classes of strongly coupled systems.

For each $d$ many dual pairs of theories with less than maximal supersymmetry are known. Less is known about such dual pairs than about the maximally supersymmetric pairs discussed in the preceding paragraph. Given the recent successes of the gauge/string correspondence in analysing the maximally supersymmetric examples discussed above, it is likely that examples of dual pairs with less supersymmetry will provide us with new insights into strongly coupled phenomena in general. One physically interesting set of such examples are the spacetimes dual to generic ${\cal N}=2$ four-dimensional superconformal YM theories coupled to matter. Building on earlier work on supersymmetric solutions with $AdS_m$ and/or $S^n$ factors~\cite{Gauntlett:2004zh},~\cite{Lin:2004nb} and~\cite{Lin:2005nh}, Gaiotto and Maldacena~\cite{Gaiotto:2009gz} (GM) have constructed an M-theory spacetime ansatz dual to general ${\cal N}=2$ four-dimensional super-conformal field theories (SCFTs). The GM solutions are implicit: given a ${\cal N}=2$ SCFT of the type discussed recently in the work of Gaiotto~\cite{Gaiotto:2009we} one has to solve the three-dimensional Toda equation
\begin{equation}
\label{toda}
\left(\partial_{x_1}^2+\partial_{x_2}^2\right)D(x_1,x_2,y)+\partial_y^2e^{D(x_1,x_2,y)}=0\,.
\end{equation}
with particular boundary conditions. General solutions of the Toda equation required for these spacetimes appear to be diffucult to find. As already observed in~\cite{Lin:2004nb}, much progress can be made when a $U(1)$ isometry exists in the $(x_1,x_2)$ plane. In this case the problem reduces to the solution of the Laplace equation~\cite{Ward:1990qt}
\begin{equation}
\frac{1}{\rho}\partial_{\rho}({\rho}\partial_{\rho}V(\rho,\eta))+\partial_{\eta}^2V(\rho,\eta)=0\,,
\label{laplace}
\end{equation}
where the transformation between $D,y,x_i$ and $V,\rho,\eta$ is given below in equation~(\ref{chvar}).
Most solutions with a $U(1)$ isometry should not be viewed as exact solutions, but rather as solutions with smearing; such solutions will be valid on length-scales longer than the typical smearing scale. The presence of the $U(1)$ isometry does however allow for a reduction of the M-theory solutions to Type IIA solutions.~\footnote{This may prove useful given that there exists an explicit formulation of perturbative IIA string-theory on a general spacetime~\cite{Duff:1987bx}.}  On the other hand, the Maldacena-Nu\~nez (MN) solution~\cite{Maldacena:2000mw} of equation~(\ref{toda}) does have a $U(1)$ isometry in the 
$(x_1,x_2)$ plane and is nevertheless an exact solution with no smearing. As we will see below, this solution will play a key role in our construction. 

In this paper we solve explicitly the Laplace equation~(\ref{laplace}) for any boundary conditions given by~\cite{Gaiotto:2009gz}, and in this way present an explicit Type IIA spacetime solution dual to any of the ${\cal N}=2$ four-dimensional SCFTs discussed in~\cite{Gaiotto:2009we}\footnote{Recently, a paper \cite{Donos:2010va} appeared on the arXiv which considers $U(1)$ reductions of the Toda equation in the context of the bubbling geometries considered in \cite{Lin:2004nb}.}. These solutions lift in an obvious way to M-theory solutions up to the smearing discussed above. In section~\ref{sec2} we review the GM solution~\cite{Gaiotto:2009gz} with and without the $U(1)$ isometry as well as the boundary conditions that correspond to particular ${\cal N}=2$ SCFTs. In particular, in section~\ref{seclambda} we discuss how the consistent boundary conditions lead to two classes of solutions. In sections~\ref{sec3} and~\ref{sec4} we present explicit solutions for these two classes of boundary conditions. In section~\ref{sec5} we discuss asymptotic properties of our solutions and we conclude in section~\ref{sec6}. Some of the technical details of the results in sections~\ref{sec3} and~\ref{sec4} are relegated to the appendix.

\section{A review of the GM solutions}
\label{sec2}

Within the context of the gauge/string correspondence, $AdS_{d+1}$ solutions of supergravity are leading candidates for the string or M-theory duals of $d$-dimensional CFTs. As such, they have received considerable attention in the literature following Maldacena's seminal paper~\cite{Maldacena:1997re}. A novel approach to finding solutions with $AdS_5$ factors and minimal (${\cal N}=1$) supersymmetry was proposed in~\cite{Gauntlett:2004zh}. Similar techniques were used by GM~\cite{Gaiotto:2009gz} to construct ${\cal N}=2$ solutions of M-theory with an $AdS_5$ factor.\footnote{The GM solutions build on earlier works~\cite{Lin:2004nb}~\cite{Lin:2005nh} which dealt with the Wick-rotated version of the problem.} In this section we summarise the main results of~\cite{Gaiotto:2009gz} as a way of setting the notation used later in the paper. We pay particular attention to solutions which exhibit an additional $U(1)$ symmetry and show how this simplifies the problem substantially. We also review the boundary conditions proposed by GM; we find in particular that there are two distinct types of boundary conditions for which we construct explicit solutions in sections~\ref{sec3} and~\ref{sec4} below.

\subsection{The GM solution}

The M-theory spacetime metric and fluxes for a background with an $AdS_5$ factor and 16 supercharges has the form~\cite{Lin:2004nb},~\cite{Gaiotto:2009gz}~\footnote{As is highlighted already in~\cite{Lin:2004nb} and~\cite{Gaiotto:2009gz}, it is not completely clear whether this is indeed the most general ansatz with these superisometries: one can imagine adding four-form flux along the $ds^2_4$ directions; some restrictions on the form of this flux were presented in~\cite{Gaiotto:2009gz} but a complete classification is still missing. A recent paper~\cite{Colgain:2010ev} addresses this issue.}
\begin{eqnarray}\label{11dmetric}
\d s_{11}^2&=&\kappa^{\frac{2}{3}}e^{2\tilde{\lambda}}\left(4\d s^2_{AdS_5}+y^2e^{-6\tilde{\lambda}}\d\widetilde{\Omega}_2^2+\d s^2_4\right)\nonumber\\
\d s^2_4&=&\frac{4}{1-y\partial_yD}(\d \chi+v)^2-\frac{\partial_yD}{y}[\d y^2+e^D(\d x^2_1+\d x^2_2)]\nonumber\\
v&=&*_2\d D\nonumber\\
G_4&=&\kappa F_2\wedge d\Omega_2\nonumber\\
F_2&=&2(\d\chi+v)\wedge \d(y^3e^{-6\tilde{\lambda}})+2y(1-y^2e^{-6\tilde{\lambda}})\d v-\partial_ye^D\d x_1\wedge \d x_2\nonumber\\
e^{-6\widetilde{\lambda}}&=&-\frac{\partial_yD}{y(1-y\partial_yD)}
\end{eqnarray}
where
$$
%0\leq\beta<2\pi	\qquad	%
0\leq\chi<2\pi	\qquad 0\leq y <y_c=N	\qquad	-\infty\leq x_i\leq\infty
$$
The function $D=D(y,x_1,x_2)$ specifies the solution and satisfies the three-dimensional Toda equation~(\ref{toda}).
The space with coordinates $(y,x_1,x_2)$ is topologically $H^2\times R^1_y$, where $H^2$ is the hyperbolic plane with coordinates $(x_1,x_2)$ and $*_2$ is the Hodge star on $H^2$. Globally, the coordinates on $H^2$ are identified by an element of a Fuchsian\footnote{A discrete subgroup of $PSL(2;\C)$.} group $\G$ to give a compact Riemann surface $\Sigma=H^2/\Gamma$.
The solution has a manifest $SU(2,2)\times SU(2)\times U(1)$ symmetry which comes from the $AdS_5$, $S^2$ and $\chi$ directions. These geometric symmetries are the same as the bosonic symmetries one finds in ${\cal N}=2$ SCFTs.

%In the polar coordinates $(r,\beta)$, the Toda equation may be written
%$$
%\frac{1}{r}\frac{\partial}{\partial r}\left(r\frac{\partial D}{\partial %r}\right)+\frac{1}{r}\frac{\partial^2D}{\partial \beta^2}+\frac{\partial^2e^D}{\partial y^2}=0
%$$

%\subsubsection{Solutions with $U(1)$ Isometry}

General solutions of the Toda equation~(\ref{toda}) are not presently known. In~\cite{Ward:1990qt} it was pointed out that solutions with a $U(1)$ symmetry in the $(x_1,x_2)$ plane are much easier to find. To see how this symmetry simplifies the problem it is useful to write the metric on the Riemann surface $\Sigma$ in polar coordinates
$$
\d s^2_{\Sigma}=\d x^2_1+\d x^2_2=\d r^2+r^2\d \beta^2\,,
$$
and consider solutions independent of $\beta$. Changing coordinates and defining $\rho=\rho(y,r)$ and $\eta=\eta(y,r)$ implicitly through a function $V(\eta,\rho)$
\begin{equation}
\label{chvar}
y=\rho\partial_{\rho}V\equiv\dot{V} \qquad  \log(r)=\partial_{\eta}V\equiv V'	\qquad	\rho^2=r^2e^D
\end{equation}
one finds that $V(\eta,\rho)$ satisfies the Laplace equation~(\ref{laplace}).
The potential $V$ can be thought of as a generating function for the coordinate transformation. The M-theory solution becomes~\cite{Lin:2004nb},~\cite{Gaiotto:2009gz}
\begin{eqnarray}\label{useful11dmetric}
\d s^2_{11}&=&\kappa^{\frac{2}{3}}\left(\frac{\dot{V}\tilde{\Delta}}{2V''}\right)^{\frac{1}{3}}\left(4\d s^2_{AdS_5} +\frac{2V''\dot{V}}{\tilde{\Delta}}\d s^2_{S^2}+ \frac{2V''}{\dot{V}}\left(\d\rho^2+ \frac{2\dot{V}}{2\dot{V}-\ddot{V}}\rho^2 \d\chi^2 +\d\eta^2\right)\right. \nonumber\\
&&\left.\qquad\qquad\qquad\qquad+\frac{2(2\dot{V}-\ddot{V})}{\dot{V}\tilde{\Delta}}\left(\d\beta+ \frac{2\dot{V}\dot{V}'}{2\dot{V}-\ddot{V}}\d\chi\right)^2 \right)\nonumber\\
\tilde{\Delta}&=&(2\dot{V}-\ddot{V})V''+(\dot{V}')^2\nonumber\\
C_3&=&2\kappa\left(-2\frac{\dot{V}^2V''}{\tilde{\Delta}}\d\chi+ \left(\frac{\dot{V}\dot{V}'}{\tilde{\Delta}}-\eta\right)\d\beta\right)\wedge \d\Omega_2
\end{eqnarray}
where $\d\Omega^2$ is the volume form on $S^2$. For solutions with an isometry in the $\beta$-direction described above, one can perform a Ka\l u\.za-Klein reduction along the $\beta$-direction with $\beta$ the M-theory circle to obtain a ten-dimensional IIA background.  It will be convenient to set $\kappa=1$ in what follows. Using the reduction ansatz\footnote{Where $\alpha=-\sqrt{\frac{d}{2(D-2)(D+d-2)}}$ and $\gamma-\frac{\alpha(D-2)}{d}$. Here $d=1$ and $D=10$.}
$$
\d s^2_{11}=e^{-2\alpha\varphi}g_{\mu\nu}\d x^{\mu}\d x^{\nu}+e^{-2\gamma\varphi}(\d\beta+A_{\mu}dx^{\mu})( \d\beta+A_{\nu}\d x^{\nu})\,,
$$
one finds the IIA string dilaton $\phi$ is
%$$
%e^{-4\varphi/3}=\left(\frac{\dot{V}\tilde{\Delta}}{2V''}\right)^{\frac{1}{3}}\frac{2(\dot{V}-\ddot{V})}%{\dot{V}\tilde{\Delta}}
%$$
%or
\begin{equation}
\label{dilaton}
e^{4\varphi}=\frac{4(2\dot{V}-\ddot{V})^3}{V''\dot{V}^2\tilde{\Delta}^2}\,,
\end{equation}
and the ten-dimensional metric is
\begin{eqnarray}\label{usefulmetric}
\d s^2_{10}&=&\left(\frac{2\dot{V}-\ddot{V}}{V''}\right)^{\frac{1}{2}}\left(4\d s^2_{AdS_5} +\frac{2V''\dot{V}}{\tilde{\Delta}}\d s^2_{S^2}+ \frac{2V''}{\dot{V}}\left(\d\rho^2+\d\eta^2\right)+ \frac{4V''}{2\dot{V}-\ddot{V}}\rho^2 \d\chi^2\right)\,. \nonumber
\end{eqnarray}
Using the reduction ansatz for the three-form $C_3=A_3+B_2\wedge\d \beta$, the IIA $B$-field and Ramond-Ramond three-form are
$$
A_1=\frac{2\dot{V}\dot{V}'}{2\dot{V}-\dot{V}}\d \chi	\qquad	A_3=-4\frac{\dot{V}^2V''}{\tilde{\Delta}}\d\chi\wedge\d\Omega_2	\qquad	B_2=2\left(\frac{\dot{V}\dot{V}'}{\tilde{\Delta}}-\eta\right)\d\Omega_2
$$
where the Ramond-Ramond one-form comes from the Ka\l u\.za-Klein reduction of the eleven-dimensional metric.

\subsection{The GM boundary conditions}

In order to find consistent spacetime solutions of the type described in the previous subsection, GM have identified certain boundary conditions on $D$, which we review presently.  
Note in~(\ref{11dmetric}) that the 2-sphere shrinks at $y=0$. The condition that the metric is well-defined at $y=0$ gives two sets of boundary conditions:
\begin{equation}
\label{todabc}
\partial_yD=0   \qquad  \text{and}  \qquad  e^D\sim \text{finite}
\end{equation}
at $y=0$. Let us translate these into boundary conditions in the $(\rho,\eta,\beta)$-space. Given that the Laplace equation~(\ref{laplace}) plays a key role in electrostatic problems, the boundary conditions in the $(\rho,\eta,\beta)$ coordinates will be reminiscent of electromagnetic boundary conditions. Recall that in these problems the line charge density can be derived using Gauss's law
$$
\oint\mathbf{E}\cdot \d\mathbf{S}=\int\mu\;\d M
$$
where the surface $S\sim\partial M$ is chosen to be a cylinder of height $L$ and coaxial with the $\eta$ axis and of radius $\rho$. $\d\mathbf{S}$ is an infinitesimal surface element with vector perpendicular to the surface and the volume element is $\d M=\rho\,\d\rho\,\d\beta\,\d\eta$. The charge volume density is $\mu=2\pi\lambda(\eta)\delta(\rho)\delta(\beta-\beta')$ so that Gauss's law is
$$
\int_{\eta=0}^L\d\eta\int_{\beta=0}^{2\pi}\d\beta\; \rho\partial_{\rho}V=\int_{\eta=0}^L\d\eta \int_{\beta=0}^{2\pi}  \d\beta\int_{\rho'=0}^{\rho}\d\rho'\rho' \lambda(\eta)\delta(\rho')\delta(\beta-\beta')
$$
where the component of the field strength perpendicular to the surface is $E_{\rho}=\partial_{\rho}V$. Comparing the $\eta$ integrands we arrive at an expression for the charge density
$$
\lambda(\eta)=\rho \partial_{\rho}V|_{\rho=0}=y(\eta,\rho=0)\,.
$$

The boundary conditions~(\ref{todabc}) translate in the $(\rho,\beta,\eta)$-space to the presence of an infinite conducting disc sitting at $\eta=0$ in the $(\rho,\beta)$-plane
$$
\partial_\rho V(\rho,\eta=0)=0\,,
$$
and a line charge density at $\rho=0$. To see this note that $y=\dot{V}=0$ is satisfied if either $\rho=0$ and/or $\partial_{\rho}V=0$ and each of these gives rise to a different set of boundary conditions. The former requires the existence of the line charge $\lambda$, while the latter will require the presence of, what we may think of as, an infinite conducting plane at $\eta=0$. We discuss these in turn below.

\subsubsection{Consistent $\lambda$ profiles}
\label{seclambda}

Regularity of the space-time imposes the boundary condition that a line charge $\lambda$ density resides at $\rho=0$. In~\cite{Gaiotto:2009gz}, it was argued that a physically sensible eleven-dimensional space-time theory imposes stringent constraints on the form of the line charge density. These constraints which we summarise below, arise from considering the quantisation of flux wrapping non-trivial cycles in the geometry
\begin{itemize}
\item The line charge density $\lambda(\eta)$ must have a piece-wise integral gradient; i.e. it must be continuous and composed of segments of the form $a_i\eta+\lambda_i$, where $a_i\in\Z$.
\item The positions of the kinks must be at integral values of $\eta$.
\item $\lambda(0)=0$.
\item The change in gradient of line segments at a kink must be integral and the gradient must decrease with successive line element; i.e. $a_i-a_{i-1}\in\Z_+$.
\end{itemize}
A kink in which the gradient changes by $k$ units gives rise to an $A_{k-1}$ singularity in the directions transverse to $AdS_5\times S^2$. If, in addition, we require that the space-time be smooth, we must impose the further constraint
\begin{itemize}
\item The change in gradient only is reduced by one at any kink; $a_i-a_{i-1}=1$.
\item The gradient of the first line segment should be 1.
\end{itemize}
An interesting observation in~\cite{Gaiotto:2009gz} was that the intercept $\lambda_i$ of the i'th line element is the total charge of the M5-branes which cause the kink at the beginning of the line segment. The integrality of the charges is ensured by the above conditions on the line charge. Generic consistent line charge densities have profiles of the form given in Figure 1.

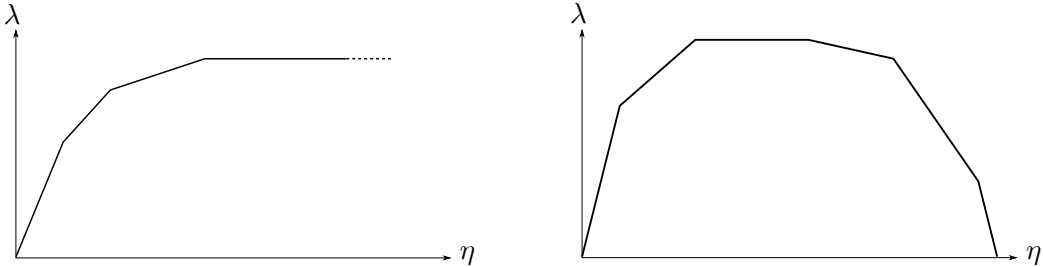
\begin{figure}[htbp] %figure placement: here, top, bottom, or page
\begin{center}
\begingroup
  \makeatletter
  \providecommand\color[2][]{%
    \errmessage{(Inkscape) Color is used for the text in Inkscape, but the package 'color.sty' is not loaded}
    \renewcommand\color[2][]{}%
  }
  \providecommand\transparent[1]{%
    \errmessage{(Inkscape) Transparency is used (non-zero) for the text in Inkscape, but the package 'transparent.sty' is not loaded}
    \renewcommand\transparent[1]{}%
  }
  \providecommand\rotatebox[2]{#2}
  \ifx\svgwidth\undefined
    \setlength{\unitlength}{401.88808594pt}
  \else
    \setlength{\unitlength}{\svgwidth}
  \fi
  \global\let\svgwidth\undefined
  \makeatother
  \begin{picture}(1,0.33036213)%
    \put(0,0){\includegraphics[width=\unitlength]{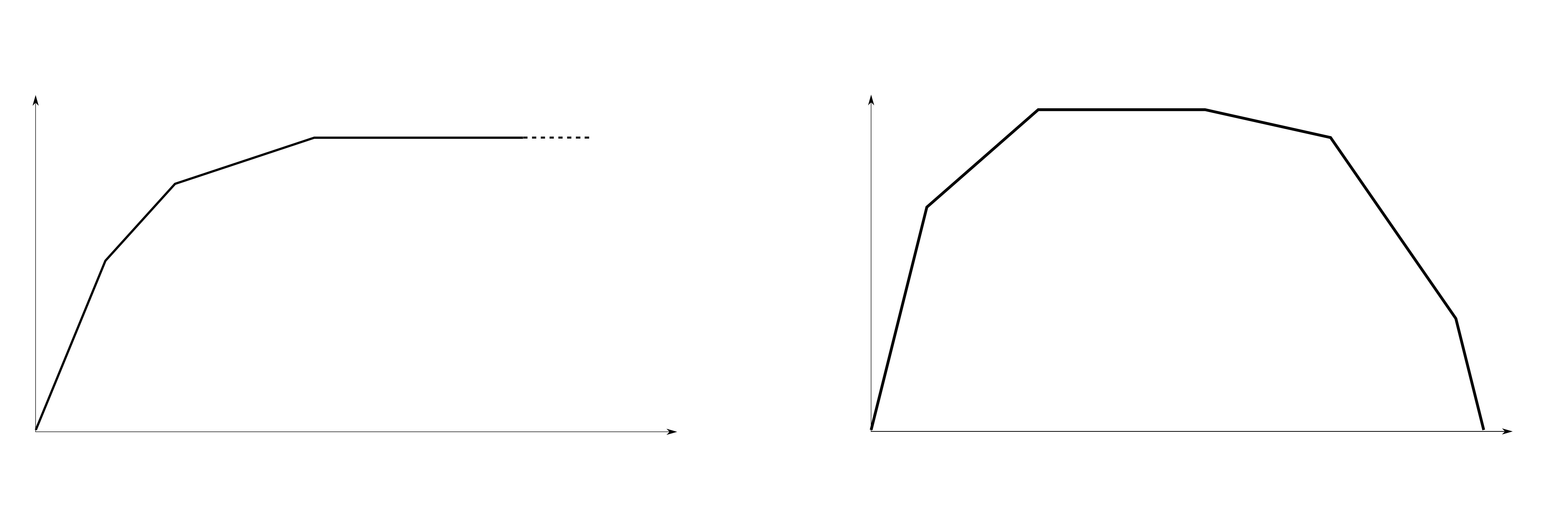}}%
    \put(0.01144526,0.2757636){\color[rgb]{0,0,0}\makebox(0,0)[lb]{\smash{$\lambda$}}}%
    \put(0.44064657,0.05357952){\color[rgb]{0,0,0}\makebox(0,0)[lb]{\smash{$\eta$}}}%
    \put(0.54433366,0.27601933){\color[rgb]{0,0,0}\makebox(0,0)[lb]{\smash{$\lambda$}}}%
    \put(0.97353493,0.05383527){\color[rgb]{0,0,0}\makebox(0,0)[lb]{\smash{$\eta$}}}%
    \end{picture}%
\endgroup
\caption{The two types of line charge; the MN-type (left) and the Uluru-type (right)}
   \end{center}
    \end{figure}

From the above restrictions on the form of $\lambda$ we see that there are two types of $\lambda$ profiles: (i) $\lambda$ intercepts the $\eta$-axis only at the origin, and (ii) $\lambda$ intercepts the $\eta$-axis at the origin and a second point $\eta=\Lambda/2>0$ for some positive real number $\Lambda$.~\footnote{This apparently awkward choice of notation will appear more natural in section~\ref{sec4}.} In the former case the physical range of $\eta$ is $\eta\ge 0$, while in the latter case it is $0\le \eta\le\Lambda/2$. Given the characteristic shape of the latter profile we will refer to these as Uluru spacetimes. The solutions we find will be quite different in the two cases and we construct them respectively in sections~\ref{sec3} and~\ref{sec4}.

\subsubsection{Infinite conducting disc}
\label{seccondplate}

We now turn to the $\partial_\rho V=0$ condition. To see how it is equivalent to an infinite conduciting disc at $\eta=0$ consider first the general condition $F(\rho,\eta)=0$, where $F$ is some function. The vanishing of this function defines a curve $\eta=\eta(\rho)$ in $(\rho,\eta)$-space. An infinitesimal variation of this gives
$$
\delta F=\partial_{\rho}F\delta\rho+\partial_{\eta}F\delta\eta=0
$$
so that the gradient of the curve is
$$
\frac{\delta\eta}{\delta\rho}=-\frac{\partial_{\rho}F}{\partial_{\eta}F}
$$
In the case at hand we have $F(\rho,\eta)=\partial_{\rho}V=0$ so that
$$
\frac{\delta\eta}{\delta\rho}=-\frac{\partial^2_{\rho}V}{\partial_{\rho}\partial_{\eta}V}
$$
What of the condition $\partial_yD=0$? We can show that $\partial_yD=2\rho^{-1}\partial_y\rho$ and furthermore, using the general formulae found in section 3.1, that the $2\times 2$ Jacobian is given by
$$
J=\rho^{-1}e^{V'}\left((\dot{V}')^2+\rho^2(V'')^2\right)
$$
The condition $\partial_yD=0$ at $y=0$ (but $\rho\neq 0$) may then be written as
$$
-\frac{1}{2}\partial_yD=\frac{\rho^{-2}\partial^2_{\eta}V}{(\partial_{\rho}\partial_{\eta}V)^2+(\partial^2_{\eta}V)^2}=0
$$
Therefore at $y=0$, the condition amounts to $\partial^2_{\eta}V=0$ or, using the Laplace equation and the fact that $\partial_{\rho}V=0$, to
$$
\partial^2_{\rho}V=0
$$
We have seen above that the boundary condition $\partial_yD=0$ corresponds to the condition $\partial^2_{\rho}V=0$. Returning to the discussion above of the gradient of the constraint surface defined by $\partial_{\rho}V=0$ we see the condition $\partial_yD=0$ requires that the constraint surface has zero gradient; $\frac{\delta\eta}{\delta\rho}=0$. In other words, the condition describes the radial field strength $\partial_{\rho}V$ vanishing on the constraint surface $\eta=$constant. We choose the constraint surface to lie at $\eta=0$. The constraint surface can be thought of as an infinite conducting disc in the $(\rho,\beta)$-plane. The reason for this interpretation of the boundary condition as a conducting disc is that the surface charge induced by a conducting disc imposes the boundary condition that the electric field has no component tangential to the conducting disc at the discs surface; i.e. that $\partial_{\rho}V$ and $\partial_{\beta}V$ vanish at $\eta=0$.

The problem of fixing the boundary conditions at the conducting plate can easily be taken care of by the method of images. The same field configuration that is caused by the semi-infinite line charge (we do not consider $\eta<0$) and conducting plane are given by extending the line charge into negative $\eta$ such that
\begin{equation*}
\lambda(\eta)\rightarrow \Lambda(\eta) = \left\{
\begin{array}{rl}
-\lambda(\eta) & \text{if }  \eta <0\\
0 & \text{if }  \eta=0\\
\lambda(\eta) & \text{if }  \eta>0
\end{array} \right.
\end{equation*}
Put simply, the condition $\partial_yD=0$ at $y=0$ is simply the requirement that the potential $V(\rho,\eta)$ must be an odd function of $\eta$. The physical region will always be restricted to $\eta\ge 0$, but below we will consider the formal solution over the whole real $\eta$ line using this method of images.

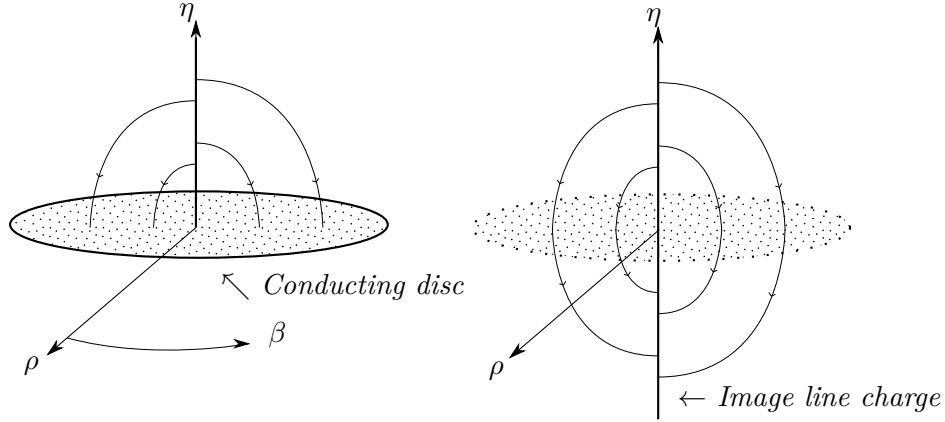
\begin{figure}[htbp] %figure placement: here, top, bottom, or page
\begin{center}
\begingroup
  \makeatletter
  \providecommand\color[2][]{%
    \errmessage{(Inkscape) Color is used for the text in Inkscape, but the package 'color.sty' is not loaded}
    \renewcommand\color[2][]{}%
  }
  \providecommand\transparent[1]{%
    \errmessage{(Inkscape) Transparency is used (non-zero) for the text in Inkscape, but the package 'transparent.sty' is not loaded}
    \renewcommand\transparent[1]{}%
  }
  \providecommand\rotatebox[2]{#2}
  \ifx\svgwidth\undefined
    \setlength{\unitlength}{400pt}
  \else
    \setlength{\unitlength}{\svgwidth}
  \fi
  \global\let\svgwidth\undefined
  \makeatother
  \begin{picture}(1,0.4)%
    \put(0,0){\includegraphics[width=\unitlength]{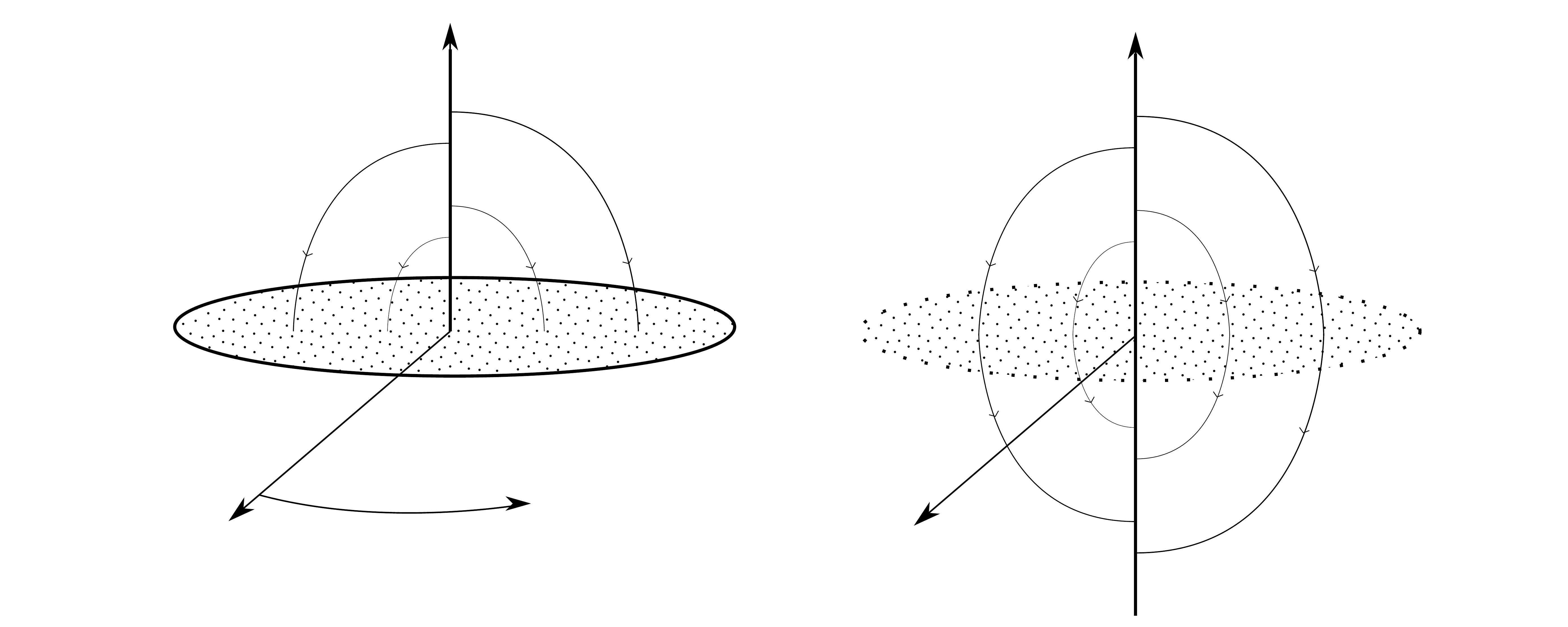}}%
    \put(0.71285713,0.38428571){\color[rgb]{0,0,0}\makebox(0,0)[lb]{\smash{$\eta$}}}%
    \put(0.56428571,0.05000002){\color[rgb]{0,0,0}\makebox(0,0)[lb]{\smash{$\rho$}}}%
    \put(0.74285714,0.01999999){\color[rgb]{0,0,0}\makebox(0,0)[lb]{\smash{\emph{$\leftarrow$ Image line charge}}}}%
    \put(0.27142856,0.38857143){\color[rgb]{0,0,0}\makebox(0,0)[lb]{\smash{$\eta$}}}%
    \put(0.12428569,0.05428572){\color[rgb]{0,0,0}\makebox(0,0)[lb]{\smash{$\rho$}}}%
    \put(0.31142854,0.12571428){\color[rgb]{0,0,0}\makebox(0,0)[lb]{\smash{\emph{$\nwarrow$ Conducting disc}}}}%
    \put(0.35571428,0.07999998){\color[rgb]{0,0,0}\makebox(0,0)[lb]{\smash{$\beta$}}}%
  \end{picture}%
\endgroup
\caption{Two ways of viewing the boundary condition}
    \end{center}
    \end{figure}

\subsection{Metric Positivity Constraints}

The metric (\ref{11dmetric}) is mostly plus (--+\dots+) for all values of $(r,y)$. If the coordinate transformation defined by the function $V(\rho,\eta)$ is a good one then the metric (\ref{usefulmetric}) should also be mostly plus for all values of $(\rho,\eta)$. In other words we must have
$$
\frac{V''\dot{V}}{\tilde{\Delta}}\geq 0 \qquad  \frac{V''}{\dot{V}}\geq 0   \qquad    \frac{V''}{2\dot{V}-\ddot{V}}\geq 0  \qquad   \frac{2\dot{V}-\ddot{V}}{\tilde{\Delta}\dot{V}}\geq 0
$$
which requires that either,
$$
V''\geq 0   \qquad  \dot{V}\geq0    \qquad  \text{or}   \qquad  V''\leq 0   \qquad  \dot{V}\leq0
$$
Using the equations of motion, we can summarise these conditions as
\begin{equation}
\ddot{V}/\dot{V}<0
\label{positivemetric}
\end{equation}
One immediate consequence of this is that $|\dot{V}|$ is a decreasing function of $\rho$.

\subsection{Dual gauge theory}

The spacetimes described in equations~(\ref{11dmetric}),~(\ref{usefulmetric}) have been conjectured by GM to be dual to ${\cal N}=2$ SCFTs. The line charge $\lambda$ can be used to read-off the generalised quiver~\cite{Gaiotto:2009we} of the dual gauge theory. Explicitly:
\begin{itemize}
\item We associate an $SU(n_i)$ gauge group with each integer value of $\eta$ (which we denote as $\eta_i$). These correspond to the circular nodes of the quiver diagram. The rank $n_i$ of the gauge group at that node is given by the value of the line charge corresponding to that point; i.e. $n_i=\lambda(\eta_i)$.
\item If there is a kink in the line charge profile, an extra $k_i$ fundamentals (hypermultiplets) are attached to the gauge group node. These are denoted by squares in the quiver diagram. The number of fundamentals $k_i$ is equal to the change in gradient at the node $n_i$. 
\end{itemize}
An example of a consistent charge profile $\lambda$ and the corresponding generalised ${\cal N}=2$ SCFT quiver is given in figure 3.

\begin{figure}[h] %figure placement: here, top, bottom, or page
\begin{center}
\begingroup
  \makeatletter
  \providecommand\color[2][]{%
    \errmessage{(Inkscape) Color is used for the text in Inkscape, but the package 'color.sty' is not loaded}
    \renewcommand\color[2][]{}%
  }
  \providecommand\transparent[1]{%
    \errmessage{(Inkscape) Transparency is used (non-zero) for the text in Inkscape, but the package 'transparent.sty' is not loaded}
    \renewcommand\transparent[1]{}%
  }
  \providecommand\rotatebox[2]{#2}
  \ifx\svgwidth\undefined
    \setlength{\unitlength}{400.88974609pt}
  \else
    \setlength{\unitlength}{\svgwidth}
  \fi
  \global\let\svgwidth\undefined
  \makeatother
  \begin{picture}(1,0.70707072)%
    \put(0,0){\includegraphics[width=\unitlength]{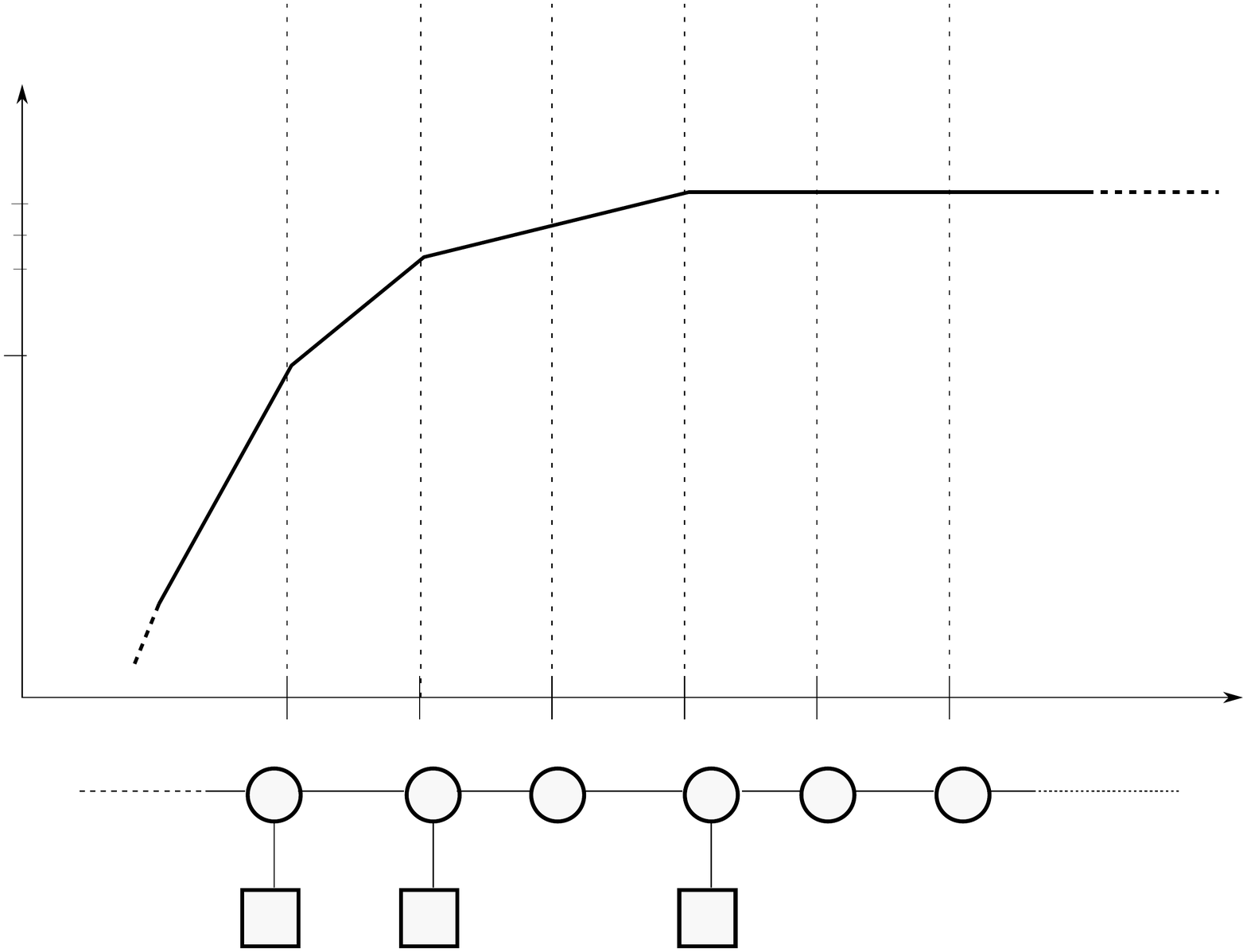}}%
    \put(0.22042748,0.03301741){\color[rgb]{0,0,0}\makebox(0,0)[lb]{\smash{$k_1$}}}%
    \put(0.53468649,0.03301741){\color[rgb]{0,0,0}\makebox(0,0)[lb]{\smash{$k_3$}}}%
    \put(0.33581414,0.03301741){\color[rgb]{0,0,0}\makebox(0,0)[lb]{\smash{$k_2$}}}%
    \put(0.23564103,0.16864165){\color[rgb]{0,0,0}\makebox(0,0)[lb]{\smash{$\eta_1$}}}%
    \put(0.1430237,0.17202832){\color[rgb]{0,0,0}\makebox(0,0)[lb]{\smash{}}}%
    \put(0.81014769,0.17281378){\color[rgb]{0,0,0}\makebox(0,0)[lb]{\smash{}}}%
    \put(0.04415496,0.66040475){\color[rgb]{0,0,0}\makebox(0,0)[lb]{\smash{$\lambda$}}}%
    \put(0.94837188,0.19581767){\color[rgb]{0,0,0}\makebox(0,0)[lb]{\smash{$\eta$}}}%
    \put(0.31981027,0.17056143){\color[rgb]{0,0,0}\makebox(0,0)[lb]{\smash{$\eta_2$}}}%
    \put(0.41550407,0.17180245){\color[rgb]{0,0,0}\makebox(0,0)[lb]{\smash{$\eta_3$}}}%
    \put(0.70057701,0.17112371){\color[rgb]{0,0,0}\makebox(0,0)[lb]{\smash{$\eta_6$}}}%
    \put(0.60487396,0.17112371){\color[rgb]{0,0,0}\makebox(0,0)[lb]{\smash{$\eta_5$}}}%
    \put(0.50984965,0.17248119){\color[rgb]{0,0,0}\makebox(0,0)[lb]{\smash{$\eta_4$}}}%
     \put(0.22127092,0.12149541){\color[rgb]{0,0,0}\makebox(0,0)[lb]{\smash{$n_1$}}}%
    \put(0.3353001,0.12149535){\color[rgb]{0,0,0}\makebox(0,0)[lb]{\smash{$n_2$}}}%
    \put(0.42489445,0.12149541){\color[rgb]{0,0,0}\makebox(0,0)[lb]{\smash{$n_3$}}}%
    \put(0.53485114,0.12081666){\color[rgb]{0,0,0}\makebox(0,0)[lb]{\smash{$n_4$}}}%
    \put(0.61765808,0.12149535){\color[rgb]{0,0,0}\makebox(0,0)[lb]{\smash{$n_4$}}}%
    \put(0.71539735,0.12149535){\color[rgb]{0,0,0}\makebox(0,0)[lb]{\smash{$n_4$}}}%
    \put(0.01714717,0.44186308){\color[rgb]{0,0,0}\makebox(0,0)[lb]{\smash{$n_1$}}}%
    \put(0.01718578,0.50362889){\color[rgb]{0,0,0}\makebox(0,0)[lb]{\smash{$n_2$}}}%
    \put(0.01708685,0.52806373){\color[rgb]{0,0,0}\makebox(0,0)[lb]{\smash{$n_3$}}}%
    \put(0.01696863,0.55249856){\color[rgb]{0,0,0}\makebox(0,0)[lb]{\smash{$n_4$}}}%
  \end{picture}%
\endgroup
\caption{There is a correspondence between the quiver diagram and the line charge. The integers $k_I\in\Z_+$ are the change in the gradient at the kink, given by $k_I=a_{I-1}-a_I$, where $a_I$ is the gradient of the $I$'th line element. Note that $\eta_I,n_I\in\Z_+$.}
  \end{center}
    \end{figure}

\section{GM spacetimes on the hyperbolic plane}
\label{sec3}

In this section we present explicit solutions for type IIA spacetimes~(\ref{usefulmetric}) by constructing the function $V(\rho,\eta)$ for a general $\lambda$ profile which intercepts the $\eta$-axis only at the origin. Such solutions can be lifted to (smeared) M-theory solutions on the hyperbolic plane. A precursor of the GM solutions of this type was the Maldacena-Nu\~nez solution~\cite{Maldacena:2000mw}, which is in fact an {\em unsmeared} M-theory solution with ${\cal N}=2$ supersymmetry and an $AdS_5$ factor. We will find that the MN solution will form a fundamental building block of the more general solutions we construct both in this and next section, and that general IIA solutions can be thought of as superpositions of suitably rescaled MN solutions. 

In section~\ref{sec31} we present the general formul{\ae}  needed for the change of variables~(\ref{chvar}). Given the central role of the MN solution in the subsequent analysis, and by way of presenting an explicit example, we review the MN solution in section~\ref{sec32} and write it in terms of $\rho,\eta$ and $V$. Following this, in section~\ref{sec33}, we show how to construct IIA solutions for a general $\lambda$ profile with a single intercept along the $\eta$-axis by superposing MN solutions. 

\subsection{Changing variables}
\label{sec31}

The change of coordinates from $y$ and $r$ to $\rho$ and $\eta$ as given in~(\ref{chvar}) is implicit and what is more varies from solution to solution. Explicit solutions for the potential $V$ may be found from the function $\rho(r,y)$ by the following general method. The starting point are the functions $\rho=\rho(r,y)$ and $\ln(r)=V'$. Differentiating $\rho$ with respect to $\eta$ and $\rho$ gives
\begin{eqnarray}\label{1}
0&=&\partial_r\rho\,\partial_{\eta}r+\partial_y\rho\,\partial_{\eta}y\nonumber\\
1&=&\partial_r\rho\,\partial_{\rho}r+\partial_y\rho\,\partial_{\rho}y
\end{eqnarray}
Differentiating $y=\dot{V}$ with respect to $\eta$, then applying the operator $\rho\partial_{\rho}$ to  $\ln(r)=V'$ and equating these two expressions for $\dot{V}'$ gives
\begin{equation}\label{3}
r\partial_{\eta}y=\rho\partial_{\rho}r
\end{equation}
Similarly appying the operator $\rho\partial_{\rho}$ to $y=\dot{V}$, then differentiating $\ln(r)=V'$ with respect to $\eta$ and finally inserting the resulting expressions into the equation of motion $\ddot{V}+\rho^2V''=0$ gives
\begin{equation}\label{2}
r\partial_{\rho}y+\rho\partial_{\eta}r=0
\end{equation}
The four equations in (\ref{1}), (\ref{3}), (\ref{2}) can be solved simultaneously to give
$$
\partial_{\rho}r=\frac{r^2(\partial_r\rho)}{{\cal M}} \qquad  \partial_{\rho}y=\frac{\rho^2(\partial_y\rho)}{{\cal M}}  \qquad  \partial_{\eta}r=-\frac{r\rho(\partial_y\rho)}{{\cal M}}  \qquad  \partial_{\eta}y=\frac{r\rho(\partial_r\rho)}{{\cal M}}
$$
where
$$
{\cal M}=r^2(\partial_r\rho)^2+\rho^2(\partial_y\rho)^2
$$
It is useful to write
$$
\left(
  \begin{array}{c}
    \d r \\
    \d y \\
  \end{array}
\right)=\left(
          \begin{array}{cc}
            \partial_{\rho}r & \partial_{\eta}r \\
            \partial_{\rho}y & \partial_{\eta}y \\
          \end{array}
        \right)\left(
  \begin{array}{c}
    \d\rho \\
    \d\eta \\
  \end{array}
\right)
$$
One may then invert the square matrix to get
$$
\left(
  \begin{array}{c}
    \d\rho \\
    \d\eta \\
  \end{array}
\right)=J^{-1}\left(
          \begin{array}{cc}
            \partial_{\eta}y & -\partial_{\eta}r \\
            -\partial_{\rho}y & \partial_{\rho}r \\
          \end{array}
        \right)\left(
  \begin{array}{c}
    \d r \\
    \d y \\
  \end{array}
\right)
$$
where $J=r\rho/{\cal M}$ is the $2\times 2$ determinant. Given an explicit expression for $D(r,y)$ and therefore an explicit expression for $\rho$ in terms of $r$ and $y$, we now find expressions for $\eta$ and $V$ by integrating
$$
\d\eta=-\frac{\rho}{r}(\partial_y\rho) \d r+\frac{r}{\rho}(\partial_r\rho) \d y
$$
and
$$
\d V=\left(\frac{y}{\rho}(\partial_r\rho)-\ln(r)\frac{\rho}{r}(\partial_y\rho)\right)\d r +\left(\frac{y}{\rho}(\partial_y\rho)+\ln(r)\frac{r}{\rho}(\partial_r\rho)\right)\d y
$$
respectively.

\subsection{The Maldacena-Nunez solution}
\label{sec32}

\begin{figure}[htbp] %figure placement: here, top, bottom, or page
\begin{center}
\begingroup
  \makeatletter
  \providecommand\color[2][]{%
    \errmessage{(Inkscape) Color is used for the text in Inkscape, but the package 'color.sty' is not loaded}
    \renewcommand\color[2][]{}%
  }
  \providecommand\transparent[1]{%
    \errmessage{(Inkscape) Transparency is used (non-zero) for the text in Inkscape, but the package 'transparent.sty' is not loaded}
    \renewcommand\transparent[1]{}%
  }
  \providecommand\rotatebox[2]{#2}
  \ifx\svgwidth\undefined
    \setlength{\unitlength}{241.88974609pt}
  \else
    \setlength{\unitlength}{\svgwidth}
  \fi
  \global\let\svgwidth\undefined
  \makeatother
  \begin{picture}(1,0.70707072)%
    \put(0,0){\includegraphics[width=\unitlength]{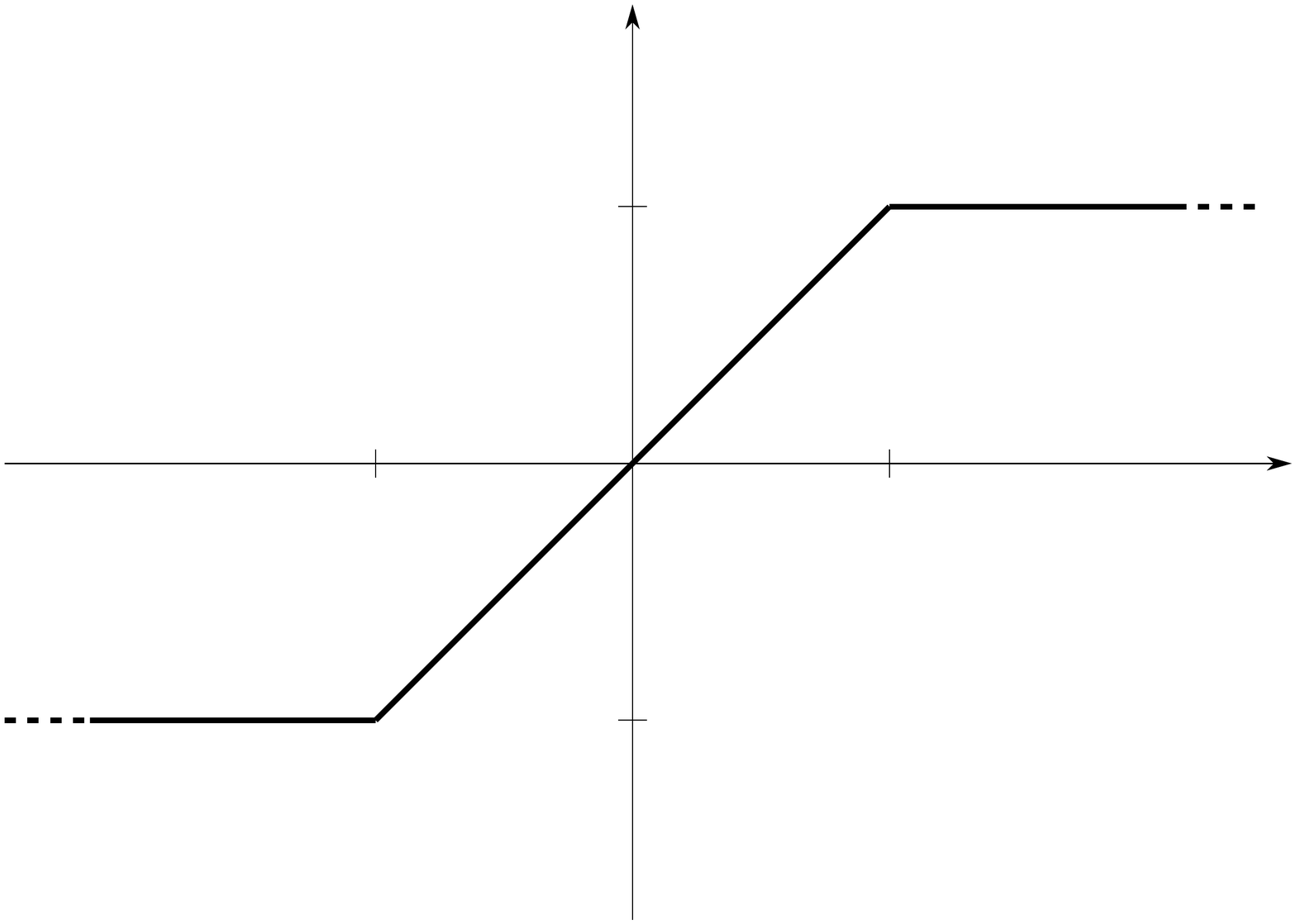}}%
    \put(0.938773,0.36091885){\color[rgb]{0,0,0}\makebox(0,0)[lb]{\smash{$\eta$}}}%
    \put(0.45114857,0.68152231){\color[rgb]{0,0,0}\makebox(0,0)[lb]{\smash{$\lambda$}}}%
    \put(0.63666294,0.31358025){\color[rgb]{0,0,0}\makebox(0,0)[lb]{\smash{$N$}}}%
    \put(0.40315404,0.53177938){\color[rgb]{0,0,0}\makebox(0,0)[lb]{\smash{$N$}}}%
    \put(0.2556463,0.31676389){\color[rgb]{0,0,0}\makebox(0,0)[lb]{\smash{$-N$}}}%
    \put(0.39083316,0.19197812){\color[rgb]{0,0,0}\makebox(0,0)[lb]{\smash{$-N$}}}%
  \end{picture}%
\endgroup
\caption{The MN line charge density profile}\label{lambdamnfig}
\end{center}
    \end{figure}

As we will see below the MN solution plays a crucial role in our construction. In this sub-section we review it paying particular attention to its description in terms of the $\rho,\eta$ and $V$ variables. The geometry describes the IR fixed point of an M5 brane wrapping the Riemann surface $\Sigma_2$ and should be viewed as a suitable quotient by a Fuchsian group acting on $r\,,\,\beta$. The eleven-dimensional metric is
\begin{eqnarray}
\d s_{11}^2&=&(\pi Nl_p^3)^{\frac{2}{3}}\frac{W^{\frac{1}{3}}}{2}\left(4\d s_{AdS_5}^2 +\frac{8}{(1-r^2)^2}(\d r^2+r^2\d\beta^2)+2\d\theta^2\right.\nonumber\\
&&\left.+\frac{2}{W}\cos^2\theta(\d\psi^2+\sin^2\theta \d\phi^2) +\frac{4}{W}\sin^2\theta\left(\d\chi+\frac{2r^2}{(1-r^2)^2}\d\beta\right)^2\right)
\end{eqnarray}
where $W=1+\cos^2\theta$.
This can be found from the general solution (\ref{11dmetric}) by the following solution of the Toda equation
$$
e^D=\frac{4}{(1-r^2)^2}(N^2-y^2)\,.
$$
The change of coordinates to the Laplace equation variables takes the form
$$
\rho(r,y)=\frac{2r}{(1-r^2)}\sqrt{N^2-y^2}\,,\qquad \qquad
\eta(r,y)=y\left(\frac{1+r^2}{1-r^2}\right)\,,
$$
with the potential given by
\begin{eqnarray}
\label{vmnsoln}
2V_{\mbox{\scriptsize MN}}(\rho\,,\,\eta\,;\,N)&=&\sqrt{\rho^2+(N+\eta)^2}-(N+\eta)\sinh^{-1}\left( \frac{N+\eta}{\rho}\right)
\nonumber \\
&&
-\sqrt{\rho^2+(N-\eta)^2}+(N-\eta) \sinh^{-1} \left(\frac{N-\eta}{\rho}\right)\,.
\end{eqnarray}
The line charge density for this solution is
\begin{equation}
\label{lambdamn}
\lambda_{\mbox{\scriptsize MN}}(\eta;N)=\left|\eta+N\right|/2-\left|\eta+N\right|/2=\left\{\begin{array}{rl}
\eta & \text{if } 0\leq \eta \leq 1\\
N & \text{if } \eta \geq 1\\
\end{array} \right.
\end{equation}
A plot of this profile, together with the $\eta\le 0$ region obtained by the method of images (see section~\ref{seccondplate}), is given in figure~\ref{lambdamnfig}

The above solution gives
$$
\dot{V}_{MN}(\rho\,,\,\eta\,;\,N)=\frac{1}{2}\sqrt{\rho^2+(N+\eta)^2}-\frac{1}{2}\sqrt{\rho^2+(N-\eta)^2}
$$
which is shown in figure 5. We see clearly that the solution gives the correct line charge boundary condition at $\rho=0$. Furthermore, we see that $|\dot{V}|$ decreases as $\rho$ becomes large, as required by the metric positivity conditions discussed in section 2.3.

\begin{center}
\begin{figure}[h] %figure placement: here, top, bottom, or page
       \centering
       \includegraphics[width=7cm]{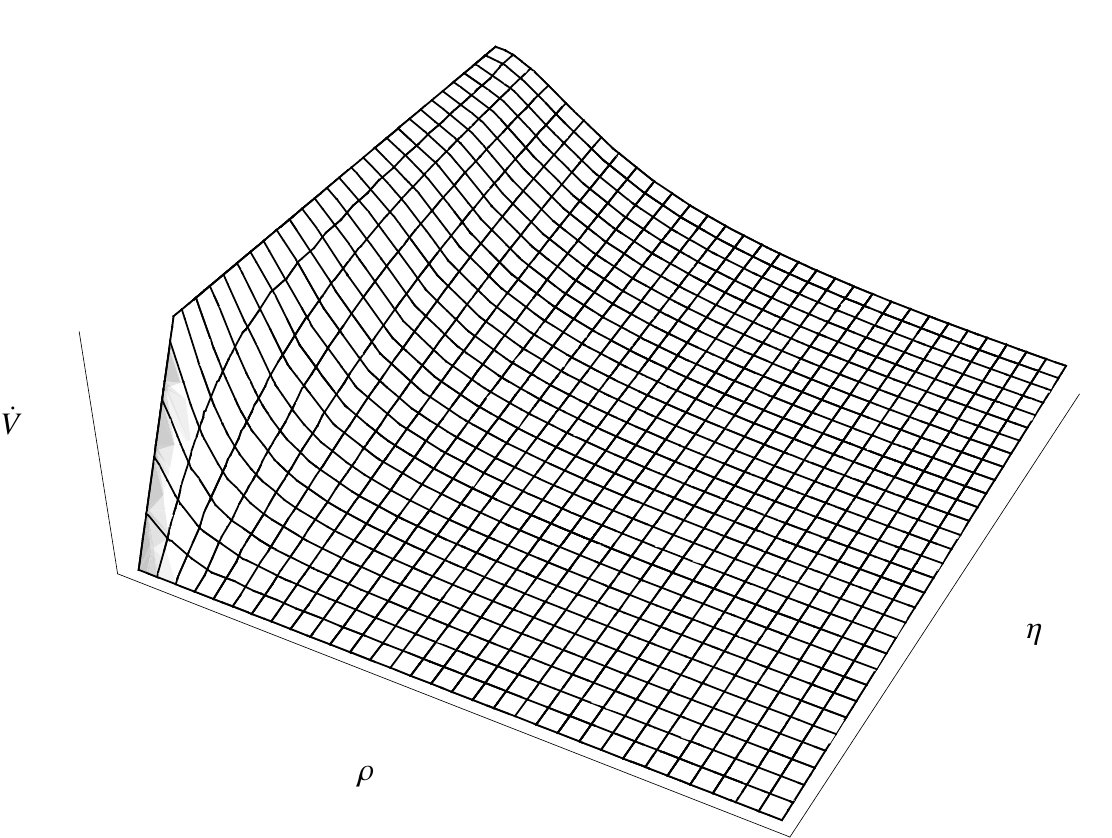}  %omit the extension
       \caption{$\dot{V}$ for the MN solution}
       \label{fig:example}
    \end{figure}
\end{center}

\subsection{General Non-periodic  Solution}
\label{sec33}

We now turn to finding $V$ for a general charge profile $\lambda$ which intercepts the $\eta$-axis only at the origin. The most general such profile that leads to a consistent spacetime is a sequence of $n$ line segments 
each of the form $\lambda_i=a_i\eta+\text{sgn}(\eta)q_i$ for constant $a_i$ and $q_i$. The charge density is thus
\begin{equation}
\label{genlambda}
\lambda(\eta) = \left\{
\begin{array}{rl}
a_0\eta & \text{if } |\eta| \leq m_1\\
a_1\eta+\text{sgn}(\eta)q_1 & \text{if } m_1\leq |\eta| \leq m_2\\
\vdots\\
a_n\eta+\text{sgn}(\eta)q_n & \text{if } m_n \leq |\eta|
\end{array} \right.
\end{equation}
where we take the line segment $\lambda_i$ to lie between points $\eta=m_i$ and $m_{i+1}$. 

To solve this problem we note two facts. Firstly, observe that any profile of the type given in equation~(\ref{genlambda}) can be viewed as a sum of suitably re-scaled and shifted $\lambda_{\mbox{\scriptsize MN}}$ profiles~(\ref{lambdamn})
\begin{equation}
\lambda(\eta)=\sum_{k=0}^n (a_{i-1}-a_i)\lambda_{\mbox{\scriptsize MN}}(\eta;m_i)\,.
\end{equation} 
A proof of this statement based on Fourier transforms is presented in the appendix. Secondly, we note that the Laplace equation~(\ref{laplace}) is linear, and so the solution satisfying the boundary condition $\dot{V}(\rho=0,\eta)=\lambda(\eta)$ may be written as
$$
V(\rho,\eta)=\sum_{i=0}^n(a_{i-1}-a_i)V_{\mbox{\scriptsize MN}}(\rho,\eta;m_i)
$$
where $V_{\mbox{\scriptsize MN}}(\rho,\eta;m_i)$ is given in equation~(\ref{vmnsoln}).

\section{Uluru spacetimes}
\label{sec4}

In the previous section we constructed spacetimes corresponding to general $\lambda$ profiles which intersect the $\eta$-axis only at the origin. As explained in section~\ref{seclambda}, a different type of profile are the Uluru profiles; for these, $\lambda$ intercepts the $\eta$ axis not just at $\eta=0$ but also at $\Lambda/2>0$. In this section we construct the general Uluru solution. We present three equivalent expressions for $V$ - each form is useful in understanding a different feature of the solution. Since the solution for a general Uluru profile is notationaly involved, we first present the simplest Uluru solution in section~\ref{sec41} to demonstrate the key features of these solutions. The general Uluru solution is given in section 4.2.

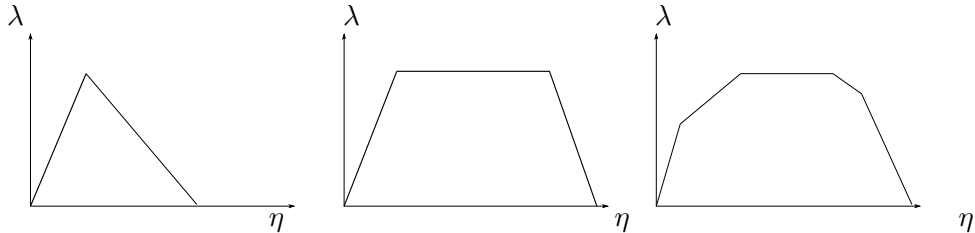
\begin{figure}[htbp] %figure placement: here, top, bottom, or page
\begin{center}
\begingroup
  \makeatletter
  \providecommand\color[2][]{%
    \errmessage{(Inkscape) Color is used for the text in Inkscape, but the package 'color.sty' is not loaded}
    \renewcommand\color[2][]{}%
  }
  \providecommand\transparent[1]{%
    \errmessage{(Inkscape) Transparency is used (non-zero) for the text in Inkscape, but the package 'transparent.sty' is not loaded}
    \renewcommand\transparent[1]{}%
  }
  \providecommand\rotatebox[2]{#2}
  \ifx\svgwidth\undefined
    \setlength{\unitlength}{400.88974609pt}
  \else
    \setlength{\unitlength}{\svgwidth}
  \fi
  \global\let\svgwidth\undefined
  \makeatother
  \begin{picture}(1,0.28895945)%
    \put(0,0){\includegraphics[width=\unitlength]{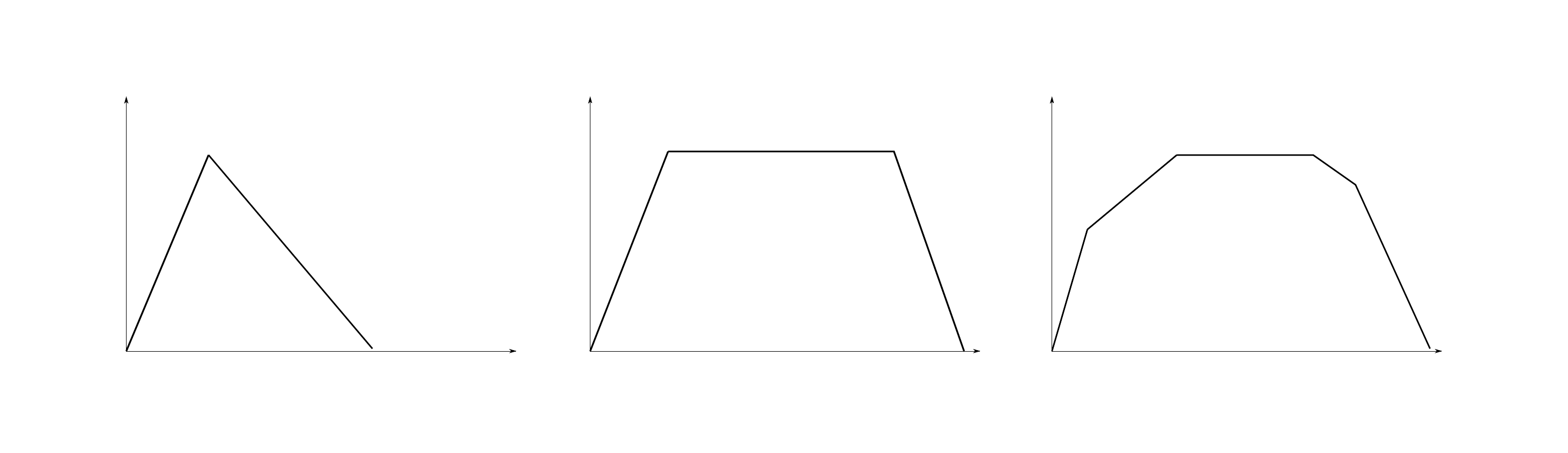}}%
    \put(0.06056341,0.02370774){\color[rgb]{0,0,0}\makebox(0,0)[lb]{\smash{}}}%
    \put(0.05861,0.23605988){\color[rgb]{0,0,0}\makebox(0,0)[lb]{\smash{$\lambda$}}}%
    \put(0.30559212,0.0457511){\color[rgb]{0,0,0}\makebox(0,0)[lb]{\smash{$\eta$}}}%
    \put(0.37649627,0.02370774){\color[rgb]{0,0,0}\makebox(0,0)[lb]{\smash{}}}%
    \put(0.37554287,0.23605988){\color[rgb]{0,0,0}\makebox(0,0)[lb]{\smash{$\lambda$}}}%
    \put(0.63152501,0.04575116){\color[rgb]{0,0,0}\makebox(0,0)[lb]{\smash{$\eta$}}}%
    \put(0.70107165,0.02370774){\color[rgb]{0,0,0}\makebox(0,0)[lb]{\smash{}}}%
    \put(0.67011826,0.23605988){\color[rgb]{0,0,0}\makebox(0,0)[lb]{\smash{$\lambda$}}}%
    \put(0.95610045,0.04575116){\color[rgb]{0,0,0}\makebox(0,0)[lb]{\smash{$\eta$}}}%
  \end{picture}%
\endgroup
\caption{Examples of Uluru-type line charge profiles}
\end{center}
    \end{figure}

\subsection{A simple Uluru spacetime}
\label{sec41}

The simplest Uluru solution has the following $\lambda$ profile
\begin{equation*}
\lambda_{\mbox{\scriptsize Uluru}}(\eta) = \left\{
\begin{array}{rl}
N\eta & \text{if } 0\leq \eta \leq 1\\
N & \text{if } 1 \leq\eta\leq K+1\\
N(K+2-\eta) & \text{if } K+1 \leq \eta\leq K+2
\end{array} \right.
\end{equation*}
This profile is interesting because the dual gauge theory is a theory with $SU(N)^K$ gauge group, $N_f=2N$ fundamental `quark' multiplets and $K-1$ scalar multiplets in the bi-fundamental. We can think of it as a distant cousin of ${\cal N}=2$ SQCD${}_K$. This theory is conformal and, in the large $N$ limit, has a large number of fields in the fundamental representation of $SU(N)$. It is also {\em not} an orbifold/orientifold or $\beta$-deformation of ${\cal N}=4$ SYM, and so provides perhaps the simplest non-trivial example of an ${\cal N}=2$ dual pair. Gauge theories of this type have recently received attention in the works~\cite{Gadde:2009dj}.

\begin{figure}[htbp] %figure placement: here, top, bottom, or page
\begin{center}
\begingroup
  \makeatletter
  \providecommand\color[2][]{%
    \errmessage{(Inkscape) Color is used for the text in Inkscape, but the package 'color.sty' is not loaded}
    \renewcommand\color[2][]{}%
  }
  \providecommand\transparent[1]{%
    \errmessage{(Inkscape) Transparency is used (non-zero) for the text in Inkscape, but the package 'transparent.sty' is not loaded}
    \renewcommand\transparent[1]{}%
  }
  \providecommand\rotatebox[2]{#2}
  \ifx\svgwidth\undefined
    \setlength{\unitlength}{250pt}
  \else
    \setlength{\unitlength}{\svgwidth}
  \fi
  \global\let\svgwidth\undefined
  \makeatother
  \begin{picture}(1,0.51701784)%
    \put(0,0){\includegraphics[width=\unitlength]{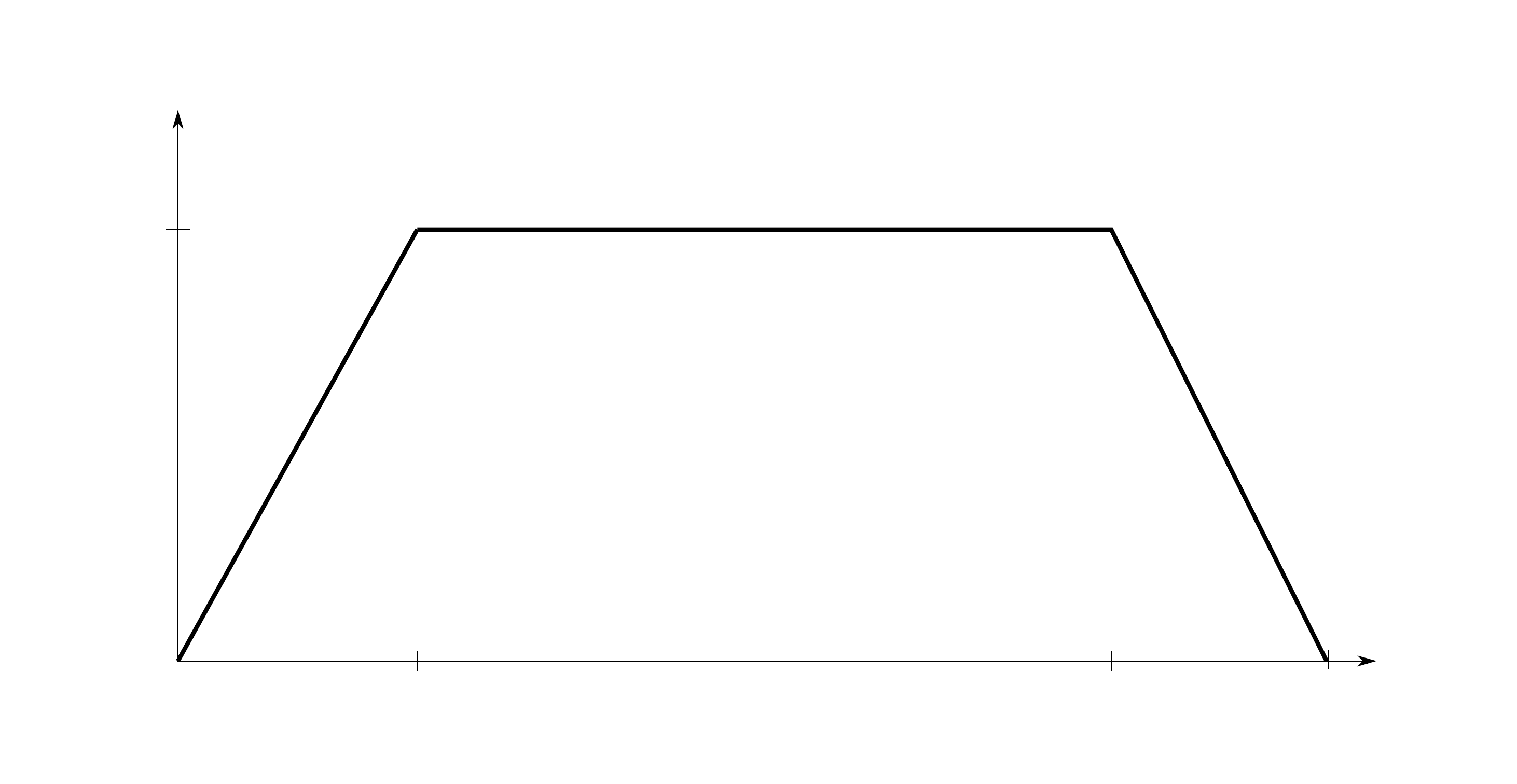}}%
    \put(0.26812397,0.02583889){\color[rgb]{0,0,0}\makebox(0,0)[lb]{\smash{$1$}}}%
    \put(0.62956426,0.02583889){\color[rgb]{0,0,0}\makebox(0,0)[lb]{\smash{$K+1$}}}%
    \put(0.11733817,0.02584521){\color[rgb]{0,0,0}\makebox(0,0)[lb]{\smash{$0$}}}%
    \put(0.79664155,0.02575361){\color[rgb]{0,0,0}\makebox(0,0)[lb]{\smash{$K+2$}}}%
    \put(0.11431052,0.46343772){\color[rgb]{0,0,0}\makebox(0,0)[lb]{\smash{$\lambda$}}}%
    \put(0.95721048,0.08052943){\color[rgb]{0,0,0}\makebox(0,0)[lb]{\smash{$\eta$}}}%
    \put(0.0621116,0.3590666){\color[rgb]{0,0,0}\makebox(0,0)[lb]{\smash{$N$}}}%
    \end{picture}%
\endgroup
\caption{A simple Uluru line charge density profile}
\end{center}
    \end{figure}
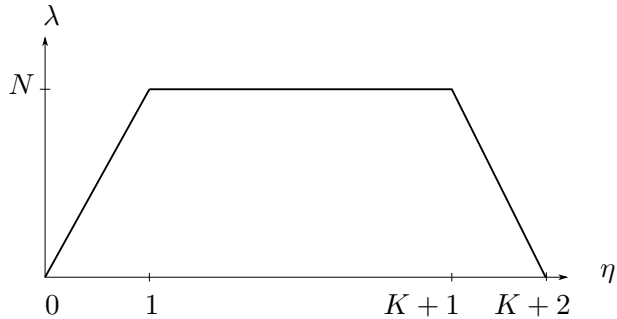

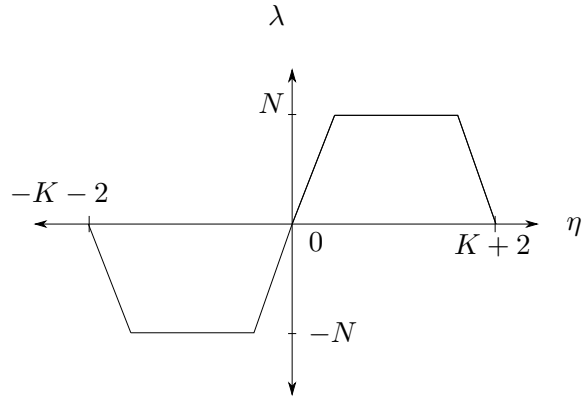
\begin{figure}[htbp] %figure placement: here, top, bottom, or page
\begin{center}
\begingroup
  \makeatletter
  \providecommand\color[2][]{%
    \errmessage{(Inkscape) Color is used for the text in Inkscape, but the package 'color.sty' is not loaded}
    \renewcommand\color[2][]{}%
  }
  \providecommand\transparent[1]{%
    \errmessage{(Inkscape) Transparency is used (non-zero) for the text in Inkscape, but the package 'transparent.sty' is not loaded}
    \renewcommand\transparent[1]{}%
  }
  \providecommand\rotatebox[2]{#2}
  \ifx\svgwidth\undefined
    \setlength{\unitlength}{400pt}
  \else
    \setlength{\unitlength}{\svgwidth}
  \fi
  \global\let\svgwidth\undefined
  \makeatother
  \begin{picture}(1,0.38705888)%
    \put(0,0){\includegraphics[width=\unitlength]{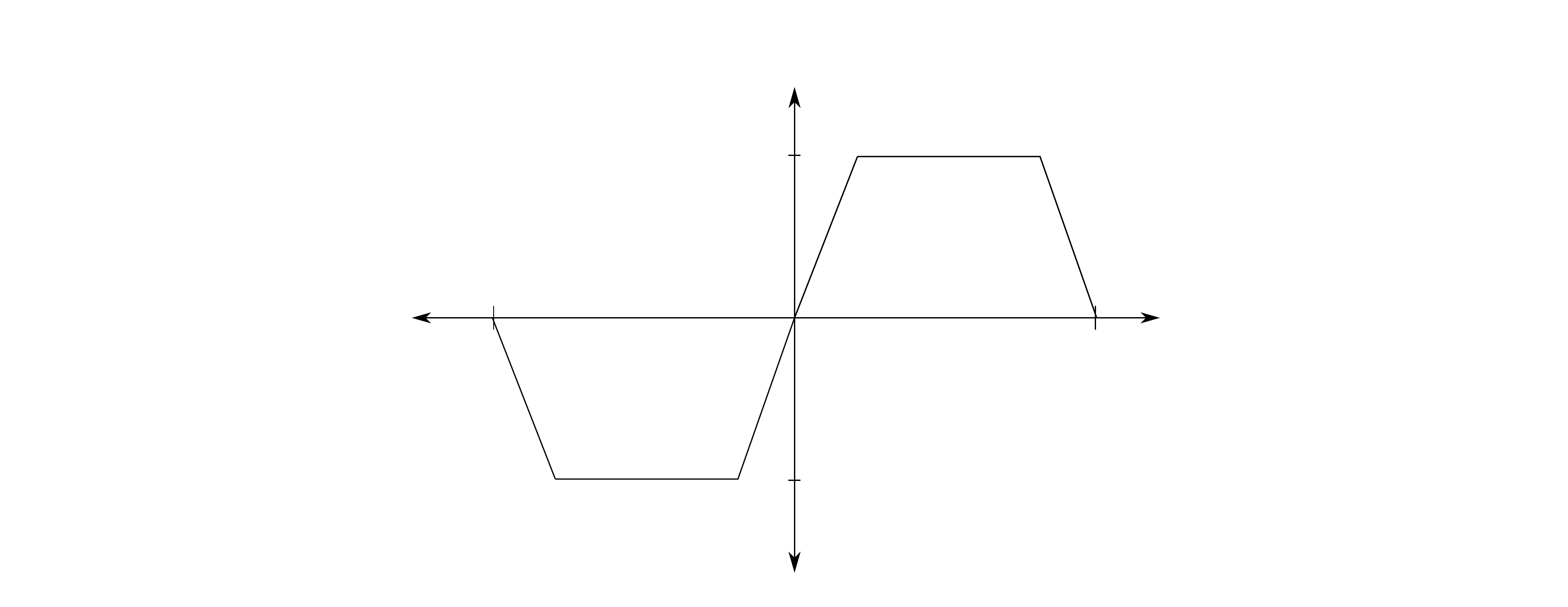}}%
    \put(0.23969673,0.20485668){\color[rgb]{0,0,0}\makebox(0,0)[lb]{\smash{$-K-2$}}}%
    \put(0.65992312,0.15599181){\color[rgb]{0,0,0}\makebox(0,0)[lb]{\smash{$K+2$}}}%
    \put(0.52238645,0.15909437){\color[rgb]{0,0,0}\makebox(0,0)[lb]{\smash{$0$}}}%
    \put(0.47507286,0.28901275){\color[rgb]{0,0,0}\makebox(0,0)[lb]{\smash{$N$}}}%
    \put(0.52212906,0.07294452){\color[rgb]{0,0,0}\makebox(0,0)[lb]{\smash{$-N$}}}%
    \put(0.4839946,0.37230353){\color[rgb]{0,0,0}\makebox(0,0)[lb]{\smash{$\lambda$}}}%
    \put(0.76632476,0.17839541){\color[rgb]{0,0,0}\makebox(0,0)[lb]{\smash{$\eta$}}}%
  \end{picture}%
\endgroup
\caption{A simple Uluru-type line charge with its image line charge}\label{ImageUluru}
\end{center}
    \end{figure}

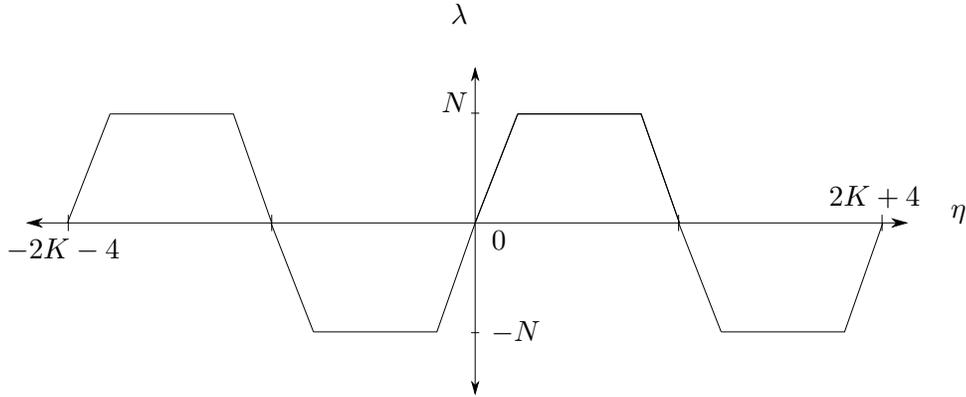
\begin{figure}[htbp] %figure placement: here, top, bottom, or page
\begin{center}
\begingroup
  \makeatletter
  \providecommand\color[2][]{%
    \errmessage{(Inkscape) Color is used for the text in Inkscape, but the package 'color.sty' is not loaded}
    \renewcommand\color[2][]{}%
  }
  \providecommand\transparent[1]{%
    \errmessage{(Inkscape) Transparency is used (non-zero) for the text in Inkscape, but the package 'transparent.sty' is not loaded}
    \renewcommand\transparent[1]{}%
  }
  \providecommand\rotatebox[2]{#2}
  \ifx\svgwidth\undefined
    \setlength{\unitlength}{400.88798828pt}
  \else
    \setlength{\unitlength}{\svgwidth}
  \fi
  \global\let\svgwidth\undefined
  \makeatother
  \begin{picture}(1,0.38705888)%
    \put(0,0){\includegraphics[width=\unitlength]{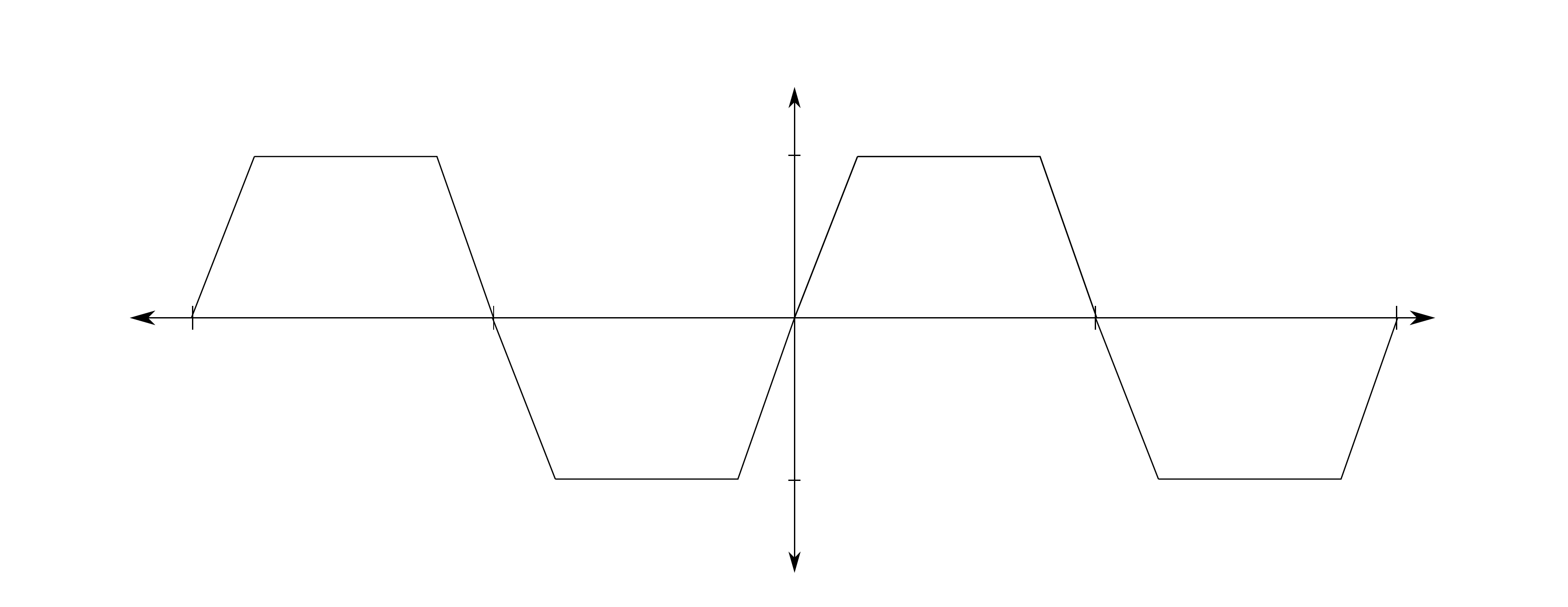}}%
    \put(0.06440381,0.15056239){\color[rgb]{0,0,0}\makebox(0,0)[lb]{\smash{$-2K-4$}}}%
    \put(0.84072865,0.20097849){\color[rgb]{0,0,0}\makebox(0,0)[lb]{\smash{$2K+4$}}}%
    \put(0.52238645,0.15909434){\color[rgb]{0,0,0}\makebox(0,0)[lb]{\smash{$0$}}}%
    \put(0.47507286,0.28901276){\color[rgb]{0,0,0}\makebox(0,0)[lb]{\smash{$N$}}}%
    \put(0.52212906,0.07294452){\color[rgb]{0,0,0}\makebox(0,0)[lb]{\smash{$-N$}}}%
    \put(0.4839946,0.37230353){\color[rgb]{0,0,0}\makebox(0,0)[lb]{\smash{$\lambda$}}}%
    \put(0.95557906,0.18925428){\color[rgb]{0,0,0}\makebox(0,0)[lb]{\smash{$\eta$}}}%
  \end{picture}%
\endgroup
\caption{A periodic Uluru-type line charge}\label{PeriodicUluru}
\end{center}
    \end{figure}

The boundary conditions discussed in section~(\ref{seccondplate}) apply equally to the $\eta=0$ and $\eta=K+2$ intercepts in the $\lambda$ profile. In particular, the method of images implies that we can look for solutions with $\eta$ on the whole real line with a periodic $\lambda$ like in figure 9. To find $V$ in this case we first note that for a periodic $\lambda$ profile we have 
\begin{equation}
\label{simpleuluru}
\lambda_{\mbox{\scriptsize Uluru}}(\eta)=\frac{N}{2}\sum_{m=-\infty}^\infty\sum_{i=1}^3\left(|\eta+m\Lambda+\nu_i|-|\eta+m\Lambda-\nu_i|\right)\,,
\end{equation}
where $\Lambda=2K+4$, $\nu_1=1$, $\nu_2=K+1$, $\nu_3=-K-2$. The above can be seen most easily by observing that a single Uluru profile and its image as depicted in figure~\ref{ImageUluru} can be written as the $m=0$ part of the above sum; periodicity along the $\eta$ axis is then achieved by summing over non-zero $m$ as shown in figure~\ref{PeriodicUluru}.

Since the Laplace equation is linear and we know the form of $V$ for each constituent in the sum~(\ref{simpleuluru}) we can write the potential $V$ as
$$
V_{\mbox{\scriptsize Uluru}}=\frac{N}{2}\sum_{l=1}^3\sqrt{\rho^2+(\nu_l+\eta)^2}-(\nu_l+\eta)\sinh^{-1}\left( \frac{\nu_l+\eta}{\rho}\right)\,.
$$
In the spacetime metric and fluxes $V$ itself does not play a role, rather $\dot{V}$ and its derivatives $\ddot{V}$, $\dot{V'}$ and $V''=-\rho^{-2}\ddot{V}$ enter the expressions directly. For the periodic Uluru potential above we have
\begin{eqnarray}
\dot{V}_{\mbox{\scriptsize Uluru}}(\rho,\eta)=\frac{N}{2}\sum_{m=-\infty}^{\infty}\sum_{l=1}^3
\sqrt{\rho^2+(\nu_l+m\Lambda-\eta)^2}-\sqrt{\rho^2+(\nu_l-m\Lambda+\eta)^2}
\label{sumvdot}
\end{eqnarray}
This expression is conceptually useful as it shows how an Uluru solution can be 'built-up' from a periodic array of MN solutions. In order to analyse its asymptotics and convergence it will be useful to re-write it as follows
\begin{eqnarray}
\dot{V}_{\mbox{\scriptsize Uluru}}(\rho,\eta)&=&\frac{N}{4}\partial_{\rho^2}^{-1}\sum_{l=1}^3\sum_{m=-\infty}^{\infty}
(\rho^2+(\nu_l+m\Lambda-\eta)^2)^{-1/2}-(\rho^2+(\nu_l-m\Lambda+\eta)^2)^{-1/2}\nonumber \\
&=&\frac{N}{2\sqrt{\pi}}\partial_{\rho^2}^{-1}\sum_{l=1}^3\sum_{m=-\infty}^{\infty}
 \int_{0}^\infty \d x \,e^{-(\rho^2+(\nu_l+m \Lambda-\eta)^2)x^2}-e^{-(\rho^2+(\nu_l-m \Lambda+\eta)^2)x^2}\nonumber \\
&=&\frac{N}{2\Lambda}\partial_{\rho^2}^{-1}\sum_{l=1}^3
 \int_{0}^\infty \frac{\d x}{x} \, e^{-\rho^2 x^2}\left[\theta_3(\xi^-_l,q)-\theta_3(\xi^+_l,q)\right]\nonumber \\
&=&\frac{N}{2\Lambda}\sum_{l=1}^3
 \int_{0}^\infty \frac{\d x}{x^3} \, e^{-\rho^2 x^2}\left[\theta_3(\xi^+_l,q)-\theta_3(\xi^-_l,q)\right]\nonumber \\
&=&\frac{N\Lambda}{4}\sum_{l=1}^3
 \int_{0}^\infty \d s \, e^{-\rho^2/ s\Lambda^2}\left[\theta_3(\xi^+_l,q)-\theta_3(\xi^-_l,q)\right]
\label{integralvdot}
\end{eqnarray}
where $s=(\Lambda x)^{-2}$ and
\begin{equation}
\xi^\pm_l=\frac{\pi(\nu_l\pm\eta)}{\Lambda}\,,\qquad q=e^{-(\pi/\Lambda x)^2}=e^{-\pi^2 s}\,.
\end{equation}
and $\theta_3$ is an elliptic theta function. The integral form of  $\dot{V}_{\mbox{\scriptsize Uluru}}$ given in equation~(\ref{integralvdot}) is useful to show that $\dot{V}_{\mbox{\scriptsize Uluru}}$ is convergent. To see this we can divide the integral into two integrals with $0<x<1$ and $1<x<\infty$, respectively. Upon a change of integration variables $x\rightarrow 1/x$ in the latter integral it is easy to see that both integrals are convergent. The only potential divergence comes from the $x\rightarrow\infty$ limit. In this limit $\theta_3(\xi^\pm_l,q)\rightarrow 1$ as can be seen from the infinite product expression for $\theta_3$
\begin{eqnarray}
\theta_3(z,q)&=&\prod_{m=1}^\infty (1-q^{2m})(1+2 \cos(z) q^{2m-1}+q^{4m-2})\label{th3prod}\,.
\end{eqnarray}
As a result, this potential divergence cancels between the two terms in the integral~(\ref{integralvdot}) above for each value of $l$ separately. It is also easy to see that the $l=3$ part of $\dot{V}$ in equation (\ref{integralvdot}) is zero.

Another useful form for $\dot{V}$ arises from the use of the following integral representation of the modified Bessel function of the second kind $K_1$
\begin{equation}
\int_{0}^\infty ds \,e^{- \rho^2/\Lambda^2s}\, e^{-n^2\pi^2s}=\frac{2\rho}{n\pi\Lambda}K_1\left(\omega_n\rho\right)\,,
\end{equation}
where
$$
\omega_n\equiv\frac{2\pi n}{\Lambda}\,.
$$
Inserting the infinite sum expression for $\theta_3$
\begin{equation}
\theta_3(z,q)=\sum_{n=-\infty}^\infty q^{n^2} e^{2i z n}\,,\label{th3sum}
\end{equation}
and using the above integral expression for $K_1$ we arrive at another useful form for $\dot{V}$
\begin{eqnarray}
\dot{V}(\rho,\eta)&=&\frac{N\Lambda}{4}\sum_{l=1}^2
\int_{0}^\infty ds \,e^{- \rho^2/\Lambda^2 s}\, \sum_{n=-\infty}^{\infty}e^{-n^2\pi^2 s}\left[e^{2 in\xi^+_l}- e^{2 in\xi^-_l}\right]\nonumber\\
&=&-N\Lambda\sum_{l=1}^2
\int_{0}^\infty ds \,e^{- \rho^2/\Lambda^2 s}\, \sum_{n=1}^{\infty}e^{-n^2\pi^2 s}
\sin\left(\omega_n\nu_l\right)\sin\left(\omega_n\eta\right)\nonumber \\
&=&-\frac{2N}{\pi}\sum_{l=1}^2\sum_{n=1}^{\infty}\frac{\rho}{n}K_1\left(\omega_n\rho\right)
\sin\left(\omega_n\nu_l\right)\sin\left(\omega_n\eta\right)\,.
\label{vdotk1}
\end{eqnarray}
To see how such an expression arises as a solution to the original Laplace problem one may use a separation of variables ansatz for $\dot{V}$ with the most general solution consistent with the boundary conditions at $\eta=0,\Lambda/2$. This takes the form
\begin{equation}
\dot{V}=\sum_{n=1}^\infty \rho\left\{A_n K_1(\omega_n\rho)+B_n I_1(\omega_n\rho)\right\}\sin(\omega_n\eta)\,,
\end{equation}
for some constant $A_n$ and $B_n$ and with $I_1$ being a modified Bessel function of the first kind. The positivity condition~(\ref{positivemetric}) implies, in the large $\rho$ limit, that the $B_n$ coefficients have to be zero. The $A_n$ coefficients can then be fixed using the Fourier series expansion for $\lambda$
\begin{equation}
\lambda_N(\eta)=\sum_{l=1,2}\sum_{n=1}^{\infty}\frac{2\Lambda}{n^2\pi^2}\sin(\omega_n\nu_l)\sin(\omega_n\eta)\,,
\label{sepvaransatz}
\end{equation}
and the $\rho\rightarrow 0$ expansion of $K_1$
$$
K_1(\omega_n\rho)=\frac{1}{\omega_n\rho}+{\cal O}(\rho)\,.
$$
This expression is perhaps the most natural one to write down for the solution of the Laplace equation~(\ref{laplace}) on an interval. It should also be clear now how to generalise this form of $\dot{V}$ for a general Uluru profile: one simply has to find the correct Fourier coefficients for $\lambda$ and insert them into the separation of variables ansatz~(\ref{sepvaransatz}). As we will see in the following section, the expression~(\ref{vdotk1}) is also useful in extracting the asymptotic behaviour of $\dot{V}$. We plot $\dot{V}$ for this Uluru solution in figure 10

\begin{center}
\begin{figure}[htbp] %figure placement: here, top, bottom, or page
       \centering
       \includegraphics[width=7cm]{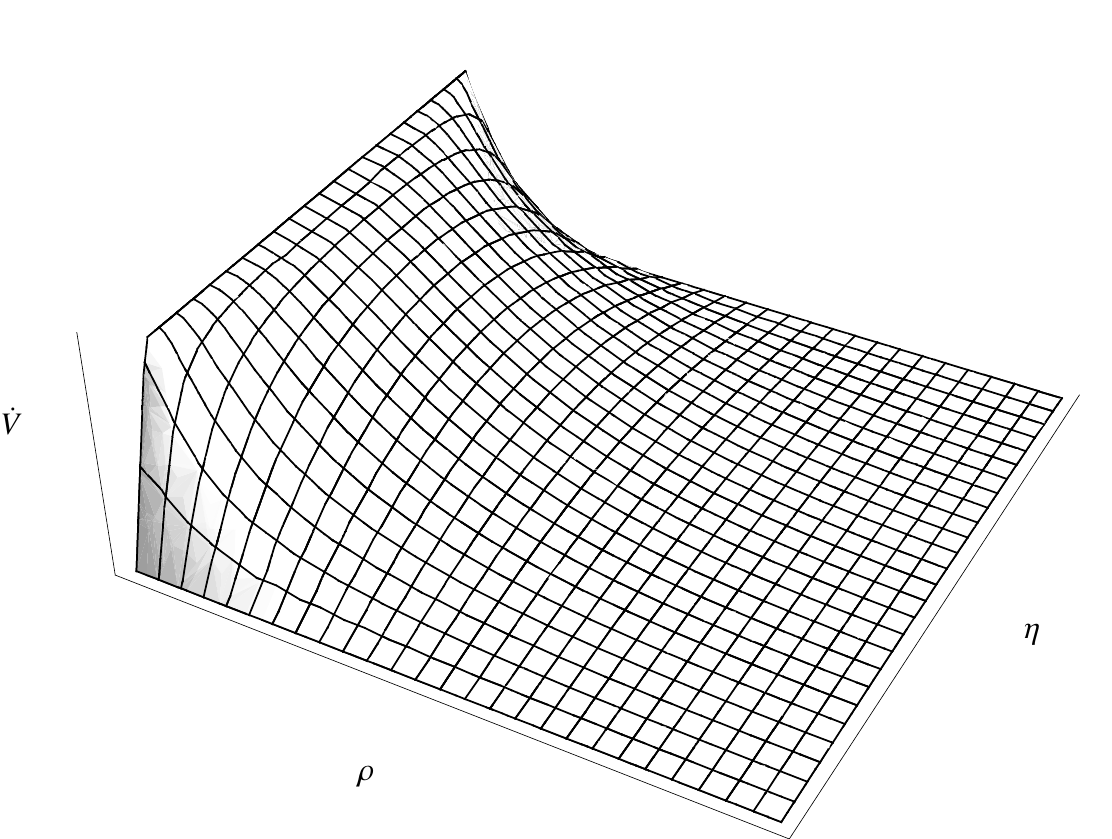}  %omit the extension
       \label{vdotulurufig}
\caption{$\dot{V}$ for the Uluru solution.}
    \end{figure}
\end{center}

\subsection{General Periodic Solution}

The considerations in the previous section generalise in a straightforward way to other Uluru-type line charge profiles. Given an Uluru-type line charge profile which has zeros at $\eta=0,\Lambda/2$, one can construct a periodic line charge of period $\Lambda$
$$
\lambda(\eta)\rightarrow \lambda_{\text{Periodic}}(\eta)\equiv \sum_{\alpha=-\infty}^{\infty}\lambda(\alpha\Lambda-\eta)
$$
and so, due to the linearity of the problem, the solution for the general periodic array is given by
$$
V(\rho,\eta)\rightarrow V_{\text{Periodic}}(\rho,\eta)\equiv \sum_{\alpha=-\infty}^{\infty}V(\rho,\alpha\Lambda-\eta)
$$
Using the result of section 3.3, that the general non-periodic line charge density may be written as a finite sum of  MN solutions, we find a corresponding statement for the general periodic solution
$$
V_{\text{Periodic}}(\rho,\eta)=-\sum_{\alpha=-\infty}^{\infty}\sum_{i=0}^{L}\left(a_i-a_{i-1}\right)V_{\text{MN}}(\rho, \alpha\Lambda-\eta;m_i)
$$
where $\{a_i\}$ are the gradients of the line elements and $\{m_i\}$ are the $\eta$-position of the kinks of the non-periodic profile, as described in~(\ref{genlambda}). There are $L$ such line elements in the range $0\leq\eta\leq\Lambda/2$ and $\Lambda=2\sum_{i=0}^Lm_i$ is the period of the line charge density profile. Following the discussion in section 4.1, $\dot{V}$ for the general periodic solution can also be written in terms of theta functions
$$
\dot{V}_{\text{Periodic}}(\rho,\eta)=\frac{\Lambda}{4}\sum_{i=0}^L\int_0^{\infty}\d s\, e^{-\rho^2/\Lambda^2s}(a_{i}-a_{i-1})\left[\theta_3(\xi^+_i,q)-\theta_3(\xi^-_i,q)\right]
$$
where
$$
\xi^{\pm}_i=\frac{\pi(m_i\pm\eta)}{\Lambda}	\qquad	q=e^{-\pi^2s}
$$
The solution may also be  written in terms of modified Bessel functions of the second kind
\begin{equation}\label{bessel}
\dot{V}_{\text{Periodic}}(\rho,\eta)=\frac{2}{\pi}\sum_{i=0}^L\sum_{n=1}^{\infty}(a_i-a_{i-1})\frac{\rho}{n}K_1(\omega_n\rho)\sin(\omega_nm_i)\sin(\omega_n\eta)
\end{equation}

\section{Behaviour at large $\rho$}
\label{sec5}

In this section we analyse the large $\rho$ behaviour of the ten-dimensional metric and dilaton for the Uluru solutions found in the previous section. We focus on the example discussed in section~\ref{sec41}, and state the corresponding (very similar) results for the general Uluru solution at the end of the section. We are interested in these asymptotics in order to establish in which regions the IIA solutions presented in the preceding sections are valid string theory backgrounds.

Consider then the periodic line charge profile given in figure 9 in the domain in which $\rho\gg K$. Our starting point is $\dot{V}$ which, in terms of an infinite sum over Bessel functions, we have shown to be given by
$$
\dot{V}=8N\sum_{n=1}^{\infty}\frac{\rho}{n\pi}K_1\left(\frac{n\pi\rho}{K+2}\right)\sin\left(\frac{n\pi}{K+2}\right)\sin\left(\frac{n\pi\eta}{K+2}\right)
$$
Assuming $n\pi\rho\gg K+2$, the modified Bessel function may be approximated by
$$
K_1(x)\approx \sqrt{\frac{\pi}{2x}}e^{-x}
$$
We shall also assume that $K\gg 2$ is large, but finite. The first term in the series dominates so that
$$
\dot{V}\approx NT\sqrt{K\rho}\;\sin\left(\frac{\pi\eta}{K+2}\right)\;e^{-\frac{\pi\rho}{K}}
\qquad	\text{where}	\qquad
T=\frac{8\sqrt{2}}{\pi}\sin\left(\frac{\pi}{K+2}\right)\neq 0
$$
From this it follows that
$$
\ddot{V}\approx-\frac{\pi NT}{\sqrt{K}}\sin\left(\frac{\pi\eta}{K+2}\right)\rho^{\frac{3}{2}}e^{-\frac{\pi\rho}{K}}
$$
We may also show that
$$
V''\approx \frac{NT\pi}{\sqrt{\rho K}}\sin\left(\frac{\pi\eta}{K+2}\right)e^{-\frac{\pi\rho}{K}}
\qquad	
\dot{V}'\approx \frac{\pi N T}{\sqrt{K}}\rho^{\frac{1}{2}}\,\cos\left(\frac{\pi\eta}{K+2}\right)e^{-\frac{\pi\rho}{K}}\,.
$$
It is useful to write these objects in terms of $\dot{V}$
$$
\ddot{V}\approx-\frac{\pi\rho}{K}\dot{V}	\qquad	V''\approx\frac{\pi}{K\rho}\dot{V}\qquad \dot{V}'\approx\frac{\pi}{K}\text{cot}\left(\frac{\pi\eta}{K+2}\right)\dot{V}
$$
As a consistency check we note that the metric positivity conditions~(\ref{positivemetric}) are satsfied in this large $\rho$ domain. The ten-dimensional IIA dilaton is given by
$$
e^{\varphi}\approx\frac{1}{NT}\sqrt{\frac{\rho}{\pi}}e^{\pi\rho/K}
$$
and the metric coefficients are
$$
\sqrt{\frac{2\dot{V}-\ddot{V}}{V''}}\approx \frac{1}{\rho^2}	\qquad	\frac{2V''\dot{V}}{\tilde{\Delta}}\approx \frac{2K}{\pi\rho}\sin^2\left(\frac{\pi\eta}{K+2}\right)	\qquad	\frac{2V''}{\dot{V}}\approx \frac{2\pi}{K\rho}	\qquad	\frac{4V''\rho^2}{2\dot{V}-\ddot{V}}\approx \text{const}
$$
For large $\rho$, the dilaton grows exponentially without limit. This growth can be slowed by taking a large $K$ or a large $N$ limit but, however large $N$ or $K$ become (provided they are finite) the dilaton will become large beyond some value of $\rho$. We conclude that, for $\rho\ll K$, the dilaton is small and the physics is approximated well by a weakly-coupled ten-dimensional string theory; however, where $\rho$ becomes much larger than $K$, the theory becomes eleven-dimensional and the perturbative string description breaks down. A similar analysis may be carried out for the general periodic case (\ref{bessel}). The key point is that $\dot{V}$ dies off at large $\rho$ - consistent with the positivity conditions of the space-time metric - and this leads to a growth in the dilaton at large $\rho$ so that, as described for the example considered above, the solution is well-described by perturbative string theory for small values of $\rho$, but M-theory is required to provide a global description of the theory. 

\section{Conclusions}
\label{sec6}

In this paper we have constructed explicit solutions of Type IIA supergravity with 16 supercharges and an $AdS_5$ factor. These solutions are conjectured to be dual to general ${\cal N}=2$ SCFTs discussed recently by Gaiotto~\cite{Gaiotto:2009we}. Our construction is completely general: given a generalised quiver describing an ${\cal N}=2$ SCFTs there is a corresponding line charge profile $\lambda$ for which we construct explicitly the corresponding spacetime metric, dilaton and fluxes. We identify two types of line-charge profiles. Firstly, there are the aperiodic profiles, of which the MN solution is the archtypical example, for which the general solution is given in section~\ref{sec3}. Secondly, there are the $\lambda$ profiles which we have called Uluru; the solutions of this type are discussed in section~\ref{sec4}. It is interesting to note that both kinds of solutions can be viewed as (finite or infinite) superpositions of MN solutions. This fits in well with our understanding of the MN solution as a dual description of a stack of M5-branes wrapping the hyperbolic plane. More general ${\cal N}=2$ SCFTs are given by some intersections of such stacks of M5-branes and our solutions validate this intuition.

The solutions we have constructed can be viewed as reductions of M-theory solutions with the GM ansatz. The reduction introduces a smearing along the M-theory circle. As a result the solutions are valid only on length scales larger than the typical smearing scale. We have seen that a generic feature of the periodic Uluru solutions are regions of spacetime for which the string coupling is large. It seems that to understand completely the dual description of these ${\cal N}=2$ SCFTs we will need to go to the full M-theory. For this one will need to find non-smeared solutions of the Toda equation~(\ref{toda}). We leave this problem to a future investigation.

\section*{Acknowledgments}

We would like to thanks Jerome Gauntlett, Dario Martelli, Carlos Nu\~nez, James Sparks, Stefan Theisen, Dan Waldram,  Lionel Mason and Fernando Alday for stimulating discussions.
BS is supported by an Advanced Research Fellowship of the EPSRC. RR is supported by EPSRC grant number EP/F016654. RR was supported by a City University Fellowship during initial stages of this work. 

\begin{appendix}

\section{The general line charge solution}

\subsection{The Maldacena-Nunez line charge}

We start with the line charge density
\begin{equation*}
\lambda(\eta) = \left\{
\begin{array}{rl}
\eta & \text{if } |\eta| \leq N\\
N\text{sgn}(\eta) & \text{if } |\eta|> N
\end{array} \right.
\end{equation*}
The charge density may then be written
$$
\lambda(\eta;N)=\int_0^{\infty}\d \zeta\,\, {\cal G}_{MN}(\zeta;N)\sin(\zeta\eta) \qquad  \text{where}    \qquad  {\cal G}_{MN}(\zeta;N)=\frac{2}{\pi}\int_0^{\infty}\d \eta\,\, \lambda(\eta;N)\sin(\zeta\eta)
$$
The expression for ${\cal G}_{MN}(k)$ splits into two parts
\begin{eqnarray}
{\cal G}_{MN}(\zeta)&=&\frac{2}{\pi}\int_0^{\infty}\d\eta\,\, \lambda(\eta)\sin(\zeta\eta)\nonumber\\
&=&\frac{2N}{\pi}\int_0^N\d\eta\,\, \eta\sin(\zeta\eta)+\frac{2N}{\pi}\int_N^{\infty}\d\eta\,\, \text{sgn}(\eta)\sin(\zeta\eta)
\end{eqnarray}
which can be evaluated to give
$$
{\cal G}_{MN}(\zeta;N)=\frac{2}{\pi \zeta^2}\sin(\zeta N)
$$

\subsection{The general non-periodic line charge}

We consider a line charge satisfying the physicality conditions summarised in section 2. Each line segment lies in the range $m_i\leq |\eta|\leq m_{i+1}$ and is of the form $\lambda_i=a_i\eta+\text{sgn}(\eta)q_i$. The requirement that $\lambda(0)=0$, imposes $q_0=m_0=0$. The full line charge density if then
$$
\lambda(\eta)=\bigcup_{i=0}^n\left(a_i\eta+\text{sgn}(\eta)q_i\right)
$$
The conditions on the line charge to give a physical spacetime theory with the correct flux quantisation conditions require $a_i,q_i,m_i\in \Z$. This line charge can be written in terms of a Fourier sine integral, as in the Maldacena-Nunez case above. The Fourier integral coefficients are
\begin{eqnarray}
{\cal G}(\zeta)&=&\frac{2}{\pi}\int_0^{\infty}\lambda(\eta)\sin(\zeta\eta)\nonumber\\
&=&\frac{2}{\pi}\sum_{i=0}^n\left[a_i\left(-\frac{\eta}{\zeta}\cos(\zeta\eta)+\frac{1}{\zeta^2}\sin(\zeta\eta)\right)+q_i\left(\frac{1}{\zeta}\left(1-\cos(\zeta\eta)\right)\right)\right]^{m_{i+1}}_{\eta=m_i}\nonumber\\
&\equiv&{\cal G}_s(\zeta)+{\cal G}_c(\zeta)\nonumber
\end{eqnarray}
where
\begin{eqnarray}
{\cal G}_s(\zeta)&=&\frac{2}{\pi \zeta^2}\sum_{i=0}^{n}a_i\left[\sin(\zeta m_{i+1})-\sin(\zeta m_i)\right]\nonumber\\
{\cal G}_c(\zeta)&=&-\frac{2}{\pi \zeta}\sum_{i=0}^{n}\ \left[\lambda_i(m_{i+1})\cos(\zeta m_{i+1})-\lambda_i(m_i)\cos(\zeta m_i)\right]\nonumber
\end{eqnarray}
and $\lambda_{\alpha}(m_{\beta})=a_{\alpha}m_{\beta}+q_{\alpha}$. In these formal sums we must remember $\lambda_{\alpha}=0$ for $\alpha$ outside of the range $0\leq\alpha\leq L$. We first focus on ${\cal G}_c(\zeta)$. Continuity of the line charge means that at the point $\eta=m_i$
$$
\lambda_{i-1}(m_i)=\lambda_{i}(m_i)
$$
means that ${\cal G}_c(\zeta)=0$. The remaining contribution to ${\cal G}(\zeta)$ is given by ${\cal G}_s(\zeta)$ which, if we change the index in the second sum, is given by
$$
{\cal G}(\zeta)=-\frac{2}{\pi \zeta^2}\sum_{i=0}^{n}\left(a_i-a_{i-1}\right)\sin(\zeta m_{i+1})
$$
However, this is simply a finite sum of Fourier integral terms for the Maldacena-Nunez line charge
$$
{\cal G}(\zeta)=-\sum_{i=0}^{n}\left(a_i-a_{i-1}\right){\cal G}_{MN}(\zeta;m_i)
$$
We see then that the general non-periodic line charge may be written as a sum of line charges for the Maldecena-Nunez solution
$$
\lambda(\eta)=-\sum_{i=0}^{n}\left(a_i-a_{i-1}\right)\lambda_{MN}(\eta;m_i)
$$
Note that since $a_i<a_{i-1}$, this is a sum of Maldacena-Nunez line charges with positive coefficients and so the metric positivity conditions on spacetimes constructed from such line charges will always be obeyed.

\subsection{General periodic line charge}

Consider a charge density profile $\lambda_{\Lambda}$ which intersects the $\eta$-axis at zero and then again at a point $\eta=\Lambda/2$, one can construct such a profile, as a finite sum of Maldacena-Nunez charge densities as described above.
$$
\lambda_{\Lambda}(\eta)=-\sum_{i=0}^{L}\left(a_i-a_{i-1}\right)\lambda_{MN}(\eta;m_i)
$$
This may then be made aperiodic, with wavelength $\Lambda$
$$
\lambda_{\Lambda}(\eta)\rightarrow \lambda^p_{\Lambda}(\eta)\equiv \sum_{\alpha=-\infty}^{\infty}\lambda_{\Lambda}(\alpha\Lambda-\eta)
$$
so that the general periodic line charge density may be written in terms of the Maldecena-Nunez line charge density
$$
\lambda^p_{\Lambda}(\eta)=-\sum_{\alpha=-\infty}^{\infty}\sum_{i=0}^{L}\left(a_i-a_{i-1}\right)\lambda_{MN}(\alpha\Lambda-\eta;m_i)
$$

\end{appendix}

\end{document}